\documentclass[11pt,a4paper]{article}
\usepackage{jheppub}
\usepackage{mathrsfs}
\usepackage{amsfonts}
\usepackage{setspace}
\usepackage{cellspace}
\usepackage{mathtools}
\usepackage{amsmath,amssymb,bm}
\usepackage[colorlinks=true,linkcolor=blue]{hyperref}
\usepackage{xcolor}
\usepackage{epsfig}
\usepackage{pifont}
\usepackage{slashed}
\usepackage{caption}
\usepackage{hhline,multirow,tabularx}  
\usepackage{dcolumn}    
\usepackage{url}        
\usepackage{braket}     
\preprint{CNF-UMD-2021}

\title{On Extraction of Twist-Two Compton Form Factors from DVCS Observables Through Harmonic Analysis}

\author[a]{Kyle~Shiells}
\author[b]{, Yuxun~Guo}
\author[a,b]{and Xiangdong~Ji}
\affiliation[a]{Center for Nuclear Femtography,\\ 1201 New York Ave., NW, Washington DC, 20005, USA}
\affiliation[b]{University of Maryland,\\ College Park, MD 20742 USA}
\emailAdd{kshiells@sura.org}
\emailAdd{yuxunguo@umd.edu}
\emailAdd{xji@umd.edu}

\abstract{We investigate the exercise of locally extracting the real and imaginary parts of the four twist-2 Compton form factors (CFFs) $\{\mathcal{H},\mathcal{E},\widetilde{\mathcal{H}},\widetilde{\mathcal{E}}\}$ which arise in the deeply virtual Compton scattering (DVCS) process $e+p\rightarrow e+p+\gamma$.  Neglecting 
dynamical higher-twist contributions, we find that there are a sufficient number of DVCS observables and degrees of freedom to extract all 8 leading quantities model-independently, exploiting the azimuthal dependence of the absolute cross sections across all possible beam and target polarizations at a common kinematical point in $\{Q,t,x_B,y\}$.  As an example, for typical JLab lab-frame kinematics, we simplify the reduced DVCS observables to their dominant terms, providing a sufficient number of equations for local determination of the twist-2 CFFs. We demonstrate the feasibility using harmonic fitting to both cross sections and beam spin asymmetries with both real and pseudo-data.}
\keywords{DVCS; GPD; Compton form factors;}
\date{\today}
\begin{document}
\maketitle

\section{Introduction}
\label{sec:1}

Hadron structure endures as a frontier field of study, requiring both cutting-edge experimental technology and theoretical descriptions.  In attempts to explain further the interesting physics behind the nucleon's properties, theoretical quantities such as generalized parton distributions (GPDs)  have been proposed and defined since around the turn of the century \cite{Muller:1994ses,Ji1997,Ji:1998pc,DIEHL2003,BELITSKY2005}.  The physics described by GPDs, for example, includes important quantities such as the intrinsic angular momentum contributions of the quarks and gluons inside the nucleon \cite{Ji1997,Ji:2020ena}.  GPDs can also allow access to the 3D tomography of the nucleon's constituents, giving unprecedented information about proton and neutron structure \cite{Burkardt2003,Belitsky:2003nz,Ji:2016djn}.

In this paper we will be focusing on some of the very relevant phenomenology questions surrounding the experimental extraction of GPDs. More specifically, we look at how one might approach the extraction of one of their natural observable counterparts: the Compton form factors (CFFs).  These particular form factors involve a convolution between a GPD (a non-perturbative nucleon matrix element) and a perturbatively-calculated Wilson coefficient, permissible by the factorization theorem \cite{Ji:1998xh, Collins1999}. CFFs naturally enter the primary amplitudes which are believed to drive exclusive scattering processes such as deeply virtual Compton scattering (DVCS) and deeply virtual meson production.  We focus this particular study to a number of physical observables associated involving the DVCS reaction $e^-+p\rightarrow e^-+p+\gamma$. 

As a matter of priority, we confine this study to the goal of model-independently extracting the twist-2 CFFs directly from DVCS observables, while reserving the inclusion of twist-3 CFFs for a future study. We also omit the target mass and finite $t$ corrections \cite{Braun:2014sta} which are kinematical twist-4 and quantum loop corrections \cite{Ji:1998xh,BELITSKY2005} that are $\alpha_S$ suppressed. Those corrections are related to the same leading-twist GPDs, but involves different Wilson coefficients that get convoluted. Therefore, they lead to a different set of CFFs that can be considered independent of the twist-2 CFFs discussed here. At present, the only known way to model-independently extract CFFs is locally in the kinematical points $(x_B,t,Q)$ at which a discrete number of measurements is performed.  Previous studies of the local extraction of CFFs \cite{Guidal2008,GuidalMoutarde2009,GuidalCLAS2010,GuidalHERMES2010,BoerGuidal2015,Kumericki2014} have been made as well as various global extraction methods \cite{Kumericki2008,Kumericki2010,Goloskokov2007,Goldstein2011}.  Global extractions typically involve some assumed model for the GPDs, allowing an interpolation of the CFFs between experimental kinematical points. One of the major challenges in extracting CFFs from exclusive measurements is the higher dimensionality of the problem, as discussed in the nice review \cite{Kumericki:2016ehc}, for example.  This has led to the somewhat recent adoption of machine learning techniques \cite{KumerickiNN2011,Moutarde:2019tqa,Cuic:2020iwt,Grigsby:2020auv} which can have the benefit of incorporating known properties of GPDs but without assuming any particular parameterized model.

In terms of previous local CFF extractions, the pioneering work of \cite{Guidal2008,GuidalMoutarde2009,GuidalCLAS2010,GuidalHERMES2010,BoerGuidal2015} has used the limited available DVCS cross section data only, together with a numerical $\chi^2$ fit to all eight twist-2 CFFs constrained to within $\pm500\%$ of their prediction from the VGG GPD model.  The general result was that the CFF $\mathcal{H}$ could in fact be constrained with a finite uncertainty (with some insight to its $t$ and $x_B$-dependence), while the remaining seven were totally unconstrained.  This study has the advantage of incorporating minimal bias and minimal approximations.  On the other hand in the subsequent work of \cite{Kumericki2014}, a more extensive set of asymmetry data was used to constrain the eight twist-2 CFFs through a linear mapping. This exercise requires somewhat strong assumptions about the relative size of various components of the total DVCS cross sections to get a linear relationship, but it allows a systematic inversion of the resulting system of equations.  The result was a finite determination of all eight CFFs with many large uncertainties, mainly due to the {\it pioneer} HERMES data itself.  

As we amass a growing ensemble of DVCS data: CLAS \cite{CLAS:2001wjj,CLAS:2006krx,CLAS:2008ahu,CLAS:2015bqi,CLAS:2015uuo}, JLab Hall A \cite{JeffersonLabHallA:2015dwe,Defurne:2017paw}, Hall B \cite{GeorgesPhD}, HERMES \cite{HERMES:2001bob,HERMES:2008abz,HERMES:2009cqe,HERMES:2010dsx,HERMES:2011bou,HERMES:2012gbh}, COMPASS \cite{COMPASS:2018pup}, ZEUS \cite{ZEUS:2008hcd}, and HERA \cite{H1:2005gdw,H1:2009wnw} as well as expect many more promised future precision data \cite{Deshpande:2016goi,AbdulKhalek:2021gbh,Anderle:2021wcy}, it is yet still a valuable exercise to further understand the prospects of a model-independent local CFF extraction.  For simplicity, we only consider 
observables in which higher-twist contributions can 
be safely neglected. The twist-2 CFFs (which we'll collectively denote by $\mathcal F$), including both their real and imaginary parts, constitute 8 unknown independent quantities which must be determined from DVCS measurements.   This beckons the question whether one can perform enough independent measurements to fully constrain all of these quantities, and even how one can guarantee the uniqueness of their solution.  Other questions include which DVCS measurements are sensitive to which CFFs, and what general strategies can be used to perform the extraction.  We answer these questions without resorting to the approximations required for a linear mapping.

We choose to adopt an approach by exploiting the azimuthal $(\phi)$ dependence in the observables: $\sigma(E_b,x_B,t,Q,\phi)$, a technique pioneered in \cite{Belitsky:2001ns,BELITSKY2005,Belitsky:2010jw}.  We will reduce our DVCS observables to the simplest possible harmonic functions in $\phi$, making explicit use of the calculated kinematics present, particularly in the cross sections and spin asymmetries we consider.  These simplifications occur because certain terms in the cross sections are kinematically suppressed (we consider a typical JLab kinematics as an example), and can be absorbed into the uncertainty estimates of the dominant terms (see for example \cite{BoerGuidal2015}).  We then demonstrate with real DVCS data that one can in fact successfully acquire enough equations which locally constrains the twist-2 CFFs, once the data is fitted to a very simple harmonic function to $\sigma(\phi)$. Fitting harmonic coefficients allows a more systematic organization of higher-twist contributions, while greatly simplifying the $\chi^2$ fit to the sought CFFs. We also offer a observable map of CFF extraction, exclusively from total DVCS cross sections.

The structure of the paper is as follows: in Section \ref{sec:2} we provide a review of the observables of the exclusive DVCS process, including a review of the azimuthal dependence and a discussion of the importance of having multiple polarizations and asymmetries.  Section \ref{sec:3} is dedicated to deriving reduced cross sections which can be used to approximate the general ones and simplify data analysis, and arrive at a practical road map for CFF extraction which can be found in Table \ref{CFFtable}. In Section \ref{sec:4} we perform a numerical analysis from available data as well as fulfill a complete CFF extraction at a chosen kinematical point to demonstrate the method of local extraction using pseudo-data.  Finally, a conclusion is given in Section \ref{sec:5}.  As a final note, many of the equations used in our analysis are left to the various Appendices, and they are referenced whenever implicitly used in the main sections.

\section{Review of DVCS Observables}
\label{sec:2}

Before we can extract the twist-2 Compton form factors (CFFs), we must have a thorough understanding of the relevant cross sections which have been and can be measured.  This section will provide the reader with an overview of many of the important observables needed to extract all of the leading CFFs, with many detailed formulas given in Appendices \ref{App:A} and \ref{App:B}.  Before we provide the details however, some historical comments are given in order to provide the context to this work.

The first DVCS cross section was computed in terms of the leading twist CFFs in \cite{JiDVCS1997}, offering a first explicit connection between the proton GPDs and experimental observables. A few years later, the DVCS cross section was recomputed by Belitsky \textit{et al.} in \cite{Belitsky:2001ns}, this time beyond the leading order GPDs, and formulas given explicitly in terms of the lab frame kinematical variables.  It was in this work also, the azimuthal dependence of the DVCS cross sections were made explicit through harmonic functions, and with it proposals on how to extract CFFs were made.  This study was then superseded by the calculation in \cite{Belitsky:2010jw} by the same authors, which included more kinematically accurate formulas, and to which we refer the collective work by BMK.

Another original calculation of the DVCS cross sections has been performed much more recently in \cite{Kriesten:2019jep} (to which we'll refer the results of as UVa).  In this study, a full inclusion of the twist-3 GPDs has been considered, utilizing also the formalism of helicity amplitudes. However, this result poses a number of substantial quantitative differences from the results of BMK, some of which are highlighted in \cite{Kriesten:2020wcx}.  

The following year yet another original calculation of the DVCS cross section was performed in \cite{Guo:2021gru}, wherein we performed a detailed study of the higher order kinematical effects accompanying the twist-2 CFFs, including the effects of light cone choices, and gauge-dependence.  Here it was found that both the BMK and UVa results could be reached once one made the appropriate coordinate choices, with the exception of an additional source of phase dependence in the interference cross section claimed by the UVa authors.  A more detailed comparison between the various results can be found in \cite{Guo:2021gru}.  

In this paper, we use our recent cross section observable results given in \cite{Guo:2021gru}, which includes a term from the Wandzura-Wilczek (WW) relation which contains additional kinematical corrections from the twist-3 CFFs, effectively improving the accuracy of the kinematics which accompanies the twist-2 CFFs. These DVCS cross sections, which are contained across Appendix \ref{App:A} and \ref{App:B} are effectively independent of light cone choice, however they do not include genuine twist-3 CFF dynamics. Although the qualitative extraction techniques for the twist-2 CFFs have not been re-invented below, the quantitative results below are indeed novel.

\subsection{Five-fold Cross Sections and Compton Form Factors Dependence}

We begin by considering the differential cross section of the reaction $e+p\rightarrow e+p+\gamma$, which comes from two amplitudes: the Compton scattering of a photon off of a struck quark in the proton (pure DVCS) and the QED-driven process of a photon being radiated from the initial and final electron beams (Bethe-Heitler).  Knowing this, we may decompose our five-fold differential cross section into 3 distinct parts: DVCS, BH and their interference ($\mathcal{I}$).  The five-fold cross section is given by
\begin{eqnarray}\label{mastersigma}
    \sigma_{\text{Tot}}(y,x_B,t,Q,\phi,\phi_S)\equiv\frac{\text{d}^5\sigma}{\text{d}x_B \text{d}Q^2\text{d}|t|\text{d}\phi \text{d}\phi_S}&=&\frac{\alpha_{\rm{EM}}^3 x_B y^2}{16\pi^2 Q^4\sqrt{1+\gamma^2}} \bigg{(}\left|\mathcal{T}_{\text{DVCS}}\right|^2+\left|\mathcal{T}_{\text{BH}}\right|^2+\mathcal{I} \bigg{)}\ \nonumber\\
    &\equiv&\sigma_{\text{DVCS}}+\sigma_{\text{BH}}+\sigma_{\mathcal{I}}\; ,
\end{eqnarray}
where $y$ is the lepton energy loss in the target rest frame and is related to the initial electron's beam energy $E_b$ via $y=Q^2/(2ME_bx_B)$ and the parameter $\gamma$ is defined by $\gamma \equiv 2M x_B/Q$. The same coordinate choices and conventions are made as in \cite{Kriesten:2019jep,Guo:2021gru}. Any future cross sections in the paper will refer to the differential one given here (unless explicitly specified otherwise), with their subscript reserved for the type: DVCS, BH or $\mathcal{I}$ and its superscript reserved for specifying the polarization of the external particles $P_{\text{beam}}P_{\text{target}}$.  For example, the polarized beam, unpolarized target interference cross section is denoted by $\sigma^{LU}_\mathcal{I}$. Since the BH cross section does not involve any CFFs, it will be frequently prescribed to be subtracted from the total cross section given the numerically well-determined elastic Dirac and Pauli nucleon form factors.  The pure DVCS and interference cross sections are given for all possible beam and target polarizations in Appendices \ref{App:A} and \ref{App:B} respectively while the unpolarized BH cross section is given in Appendix \ref{App:D}.

It is important to stress that part of the $(x_B,t,Q)$ dependence of $\sigma_{\text{Tot}}$ comes from the CFFs, and part of this dependence comes from exactly calculable kinematics.  However, the dependence on the other two kinematic variables $(y,\phi)$ comes completely from the kinematics, and is exactly calculable.  These extra two \textit{experimental degrees of freedom} can and should be fully exploited for the phenomenological extraction of the CFFs.  It is generally known now that as the beam energy increases in a fixed target photoproduction scattering reaction, the ratio of the BH to DVCS contributions decreases (see a first discussion of this in \cite{Belitsky:2001ns} and a nice graphical demonstration in \cite{Kriesten:2020wcx}).  This enhancement of CFF-sensitive components indeed justifies going to higher beam energy.  Meanwhile, a systematic organization of the azimuthal-dependence of $\sigma_{\text{Tot}}$ via harmonics has been heavily endorsed over the last 20 years, and we shall look in detail at this in Section \ref{sec:3}.

Amongst all of our possible polarization combinations covered in Appendices \ref{App:A} and \ref{App:B} together with the azimuthal behavior of transversely-polarized target cross sections discussed in Section \ref{sec:2}, we have a maximum of 8 distinct cross sections that can be measured -- and therefore a complete system of equations to our unknowns.  Some previous attempts to extract CFFs have been underdetermined  (involved less than 8 observables), and as a consequence, were accompanied with very large, if not infinite, error bars.  We stress the importance of a sufficiently constraining set of observables, which is required to determine one's CFFs reliably.

Unlike the application of this exercise to linearized asymmetries only (as was done in \cite{Kumericki2014}), here we have quadratic expressions in the CFFs due to the presence of the pure DVCS contributions.  The consequent nonlinear algebra greatly complicates how one finds a general solution, introducing sign ambiguities in the CFFs for instance. An arbitrary exclusive cross section from Eq.(\ref{mastersigma}), for example Eqs.(\ref{sigdvcsUU})+(\ref{sigIUU}), may be loosely expressed via
\begin{eqnarray}
\sigma^{P_bP_t}_{\text{Tot-BH}}\sim A(CFF)^2 + B(CFF)\;.
\end{eqnarray}
More precisely, we may represent such equations in the following way
\begin{eqnarray}\label{8by8CFFsystem}
    \langle \mathcal{F}|A^{(i)}|\mathcal{F}\rangle + \langle B^{(i)}|\mathcal{F}\rangle + C^{(i)} &=& 0 \;,
\end{eqnarray}
where $i$ represents distinct DVCS cross sections, the $C^{(i)}$ are the BH-subtracted cross sections, and $|\mathcal{F}\rangle=(\text{Re}\mathcal{H},\text{Im}\mathcal{H},\text{Re}\mathcal{E},\text{Im}\mathcal{E},\text{Re}\mathcal{\widetilde{H}},\text{Im}\mathcal{\widetilde{H}},\text{Re}\mathcal{\widetilde{E}},\text{Im}\mathcal{\widetilde{E}})^T$.  In principle, we would require at least 8 such equations to properly constrain all of the twist-2 CFFs, otherwise we might expect an infinite number of solutions, which would be realized by grossly oversized uncertainties on the extracted CFFs. Were we to write out these equations explicitly and attempt to solve for $|\mathcal{F}\rangle$, we would attain an overly-cumbersome symbolic result -- too long to provide here.  On the other hand, once numerical cross section data is provided through available $C^{(i)}$, the solution may be attained numerically at some common kinematical point $(x_B,Q^2,t,y,\phi)$. Implicit in our Appendices, approximated forms of all of the possible distinct matrices $A^{(i)}$ and vectors $|B^{(i)}\rangle$ are provided.

Within Eq. (\ref{8by8CFFsystem}), the matrices $A^{(i)}$ are both real and symmetric, the vectors $|B^{(i)}\rangle$ are real and of course the numbers (cross sections) $C^{(i)}$ are all real.  Physically, we require that any and all solutions $|\mathcal{F}\rangle$ are purely real. In any case one can think of Eq.(\ref{8by8CFFsystem}) as a type of generalized \textit{quadratic equation}, for which we would like to attain and understand the solution.  As we shall see later, some observables $C^{(i)}$ come with the convenient approximation that $A^{(i)}=0$, but some certainly cannot rely on this feature. In Section \ref{sec:4}, we look how to attain a physical solution to the CFFs prescribed by Eq.(\ref{8by8CFFsystem}).



As discussed, one can in principle extract all 8 twist-2 CFFs given enough polarized cross sections are measured in the lab.  Although other strategies to achieve this important extraction exercise exist (for example, see \cite{Kriesten:2020apm}), in this section we will look at using the harmonic ($\phi$) dependence.  As we shall see, given enough statistics over this angle, one can extract all of the twist-2 CFFs with as little as 6 polarization cases, affording one with the option of strengthening the CFF constraints further with additional observables.

Although total cross sections are prone to stray normalization factors and introduce challenging systematic errors, they are quite ideal for extracting CFFs, due to the straightforward nature of Eq. (\ref{mastersigma}), especially its azimuthal structure.  For this reason, much time will be spent looking at total cross sections in this section. The pure DVCS cross section is inherently quadratic in the CFFs and it also has the feature of including both their real and imaginary parts. Meanwhile the interference cross section is a very ideal contribution for studying CFFs owing to its more straightforward linear dependence on them, and for this reason, our analysis will be heavily dependent on an accurate knowledge of this contribution. 

\subsection{Polarization Observables}

The polarization degree of freedom for both the beam and target particles allows the possibility of multiple distinct observables which can be measured for the $e+p\rightarrow e+p+\gamma$ process.  Doing so allows us to span $i\in (UU,LU,UL,LL,UT,LT)$ in Eq.(\ref{8by8CFFsystem}), while the $UT$ and $LT$ cross sections can be further divided in to 2 sub-components, as we shall discuss below.  This will indeed ensure we have a sufficiently constrained system of equations.  Asymmetries involve specific linear combinations of polarized cross sections in a rational expression.  Consequently, one can think of cross sections (and all their polarization combinations) as the independent experimental inputs, while regarding asymmetries as non-independent experimental inputs.  We shall study asymmetries nonetheless as they are typically praised for having smaller experimental uncertainties than total cross sections.

\subsubsection{Transversely Polarized Targets}

We take here a logistical outlook on cross sections with a transversely polarized target as they have unique aspects not found in the other polarization cases, including a dependence on the azimuthal angle $\phi_S$.  We characterize transversely-polarized target observables with the target's polarization vector either being \textit{in plane} (parallel to final state hadronic plane) or being \textit{out of plane} (perpendicular to the hadronic plane). Consider the real $UT$ cross section for an arbitrary transverse polarization angle, from which we subtract the BH contribution, giving us
\begin{eqnarray}
\sigma^{UT}_{\text{Tot-BH}}(\phi,\phi_S)=\sigma^{UT,\rm{in}}_{\text{DVCS}+\mathcal{I}}\cos(\phi_S-\phi) + \sigma^{UT,\rm{out}}_{\text{DVCS}+\mathcal{I}}\sin(\phi_S-\phi)\;,
\end{eqnarray}
meaning that for an arbitrary $\phi$ and $\phi_S$, the measured $UT$ cross section will have both an \textit{in plane} and \textit{out of plane} component, each having a different dependence on the CFFs.  However, we can consider certain cases of controlling these two angles.  Two theoretically simple cases are
\begin{equation}\label{phiequalphiS}
\phi=\phi_S \;\;\Rightarrow \sin(\phi_S-\phi)=0 \;,
\end{equation}
\begin{equation}\label{phiequalphiSminus90}
\phi_S-\phi=\frac{\pi}{2} \;\;\Rightarrow \cos(\phi_S-\phi)=0\;. 
\end{equation}
This still allows the two angles to span any value, but merely fixes them with respect to each other.  These two cases would allow one to get two distinct cross sections from a $UT$ experiment.  We can also in principle do this for the $LT$ experiment, totaling 8 potential polarization cross sections.

Theoretically more complicated but perhaps easier experimentally are the following two $UT$ cross section observables
\begin{eqnarray}
\Delta\sigma^{UT}_{\text{Tot-BH},\pi}&\equiv&\sigma^{UT}_{\text{Tot-BH}}(\phi,\phi_S)-\sigma^{UT}_{\text{Tot-BH}}(\phi,\phi_S+\pi)\;,\\
\Delta\sigma^{UT}_{\text{Tot-BH},\pi/2}&\equiv&\sigma^{UT}_{\text{Tot-BH}}(\phi,\phi_S)-\sigma^{UT}_{\text{Tot-BH}}\bigg{(}\phi,\phi_S+\frac{\pi}{2}\bigg{)}\;.
\end{eqnarray}
Using trigonometric sum-of-angle identities then gives us
\begin{eqnarray}
\Delta\sigma^{UT}_{\text{Tot-BH},\pi}&=& 2\sigma^{UT,\rm{in}}_{\text{DVCS}+\mathcal{I}}\cos(\phi_S-\phi)+2\sigma^{UT,\rm{out}}_{\text{DVCS}+\mathcal{I}}\sin(\phi_S-\phi)\;,\label{delsigmaUTpi}\\
\Delta\sigma^{UT}_{\text{Tot-BH},\pi/2}&=&(\sigma^{UT,\rm{in}}_{\text{DVCS}+\mathcal{I}}-\sigma^{UT,\rm{out}}_{\text{DVCS}+\mathcal{I}})\cos(\phi_S-\phi)+(\sigma^{UT,\rm{in}}_{\text{DVCS}+\mathcal{I}}+\sigma^{UT,\rm{out}}_{\text{DVCS}+\mathcal{I}})\sin(\phi_S-\phi)\nonumber\;.\label{delsigmaUTpiby2}\\
\end{eqnarray}
For the purposes of extracting CFFs, it would be simpler if we could relate just the $UT,\rm{in}$ and $UT,\rm{out}$ cross sections to data.  Therefore we shall consider the LHS of Eq.(\ref{delsigmaUTpi}) and Eq.(\ref{delsigmaUTpiby2}) as the measurements and invert the system of equations to find the in and out cross sections via
\begin{eqnarray}
\sigma^{UT,\rm{out}}_{\text{DVCS}+\mathcal{I}} &=& \frac{1}{2}(1-\tan(\phi_S-\phi))\Delta\sigma^{UT}_{\text{Tot-BH},\pi}-\Delta\sigma^{UT}_{\text{Tot-BH},\pi/2}\;, \\
\sigma^{UT,\rm{in}}_{\text{DVCS}+\mathcal{I}} &=& \frac{1}{2}\sec(\phi_S-\phi)\Delta\sigma^{UT}_{\text{Tot-BH},\pi} -2\csc(\phi_S-\phi)\Delta\sigma^{UT}_{\text{Tot-BH},\pi/2}\;.
\end{eqnarray}
Where now the experimental measurements construct the RHS of each equation, and the LHS is then equated to the known CFF expressions which they represent, which can be found from Eqs.(\ref{sigdvcsUtout}), (\ref{sigIUTin}) \& (\ref{sigIUTout}).  The $LT$ case may be handled by an analogous argument.

\subsubsection{Asymmetries}

Since asymmetries are functions of cross sections, we regard cross sections as the independent observable degrees of freedom and hence as independent input towards CFF extraction.  We therefore regard asymmetries as a secondary means of constraining the CFFs in this paper, and consequently do not emphasize them to the same extent as we will for cross sections.
DVCS charge asymmetries have previously been measured extensively at for example, HERMES \cite{HERMES:2001bob,HERMES:2008abz,HERMES:2009cqe,HERMES:2010dsx,HERMES:2011bou,HERMES:2012gbh} and COMPASS \cite{COMPASS:2018pup}, and have indeed been used to constrain twist-2 CFFs.  Since we are choosing to restrict ourselves in this study to electron beams only, we shall look instead here at electron spin asymmetries.  Particularly, the electron beam single spin asymmetry is defined by
\begin{equation}\label{ALRoriginal}
    A_{LR}=\frac{\sum_{\Lambda}\sigma_{\text{Tot}}(h=\frac{1}{2})-\sigma_{\text{Tot}}(h=-\frac{1}{2})}{\sum_{\Lambda}\sigma_{\text{Tot}}(h=\frac{1}{2})+\sigma_{\text{Tot}}(h=-\frac{1}{2})}\;,
\end{equation}
where $h$ stands for the electron beam helicity and the sum over $\Lambda$ is over all possible target polarizations.  The numerator of $A_{LR}$ (or also denoted $A_{LU}$ in many literatures) ensures that only terms odd in $h$ will survive, whereas in the denominator only terms even in $h$ remain.  For this reason, the denominator is equivalent to the total $UU$ cross section while the numerator extracts the $LU$ interference cross section only.  Thus the beam asymmetry is given by
\begin{equation}\label{ALRsimple}
    A_{LR}=\frac{\sigma^{LU}_\mathcal{I}}{\sigma^{UU}_{\text{BH}}+\sigma^{UU}_{\text{DVCS}}+\sigma^{UU}_\mathcal{I}}\Bigg{|}_{h=\frac{1}{2}}\; ,
\end{equation}
which is a rational expression involving cross sections and its azimuthal structure is largely determined by its numerator.

This quantity requires us to know the unpolarized BH cross section, which is given explicitly in Appendix \ref{App:D} for this very purpose.  In the past, asymmetries such as $A_{LR}$ have been approximated under the assumption that the BH cross section in the denominator dominates the DVCS and interference ones.  However, this is not well-justified in many cases.  For instance in \cite{Kriesten:2020wcx}, one can see that the unpolarized interference and DVCS cross sections are at least of order 10\% of the BH cross section.  Therefore, we shall keep the full denominator and will investigate any numerically justified approximations in Section \ref{sec:5}.

Phenomenologically speaking, this sort of observable eliminates the need for exact normalization factors in the cross section and it cancels various systematic uncertainties and they are therefore ideal for CFF extraction.  On the other hand asymmetries have a more complicated dependence on both our CFFs and a more complicated azimuthal dependence.  We shall turn to sort out its azimuthal dependence in section \ref{sec:3} and then look at practical ways of extracting the CFFs from it in Section \ref{sec:4}.  We also stress that once one expresses the other spin asymmetries explicitly in terms of cross sections as in Eq.(\ref{ALRsimple}), all of the information we will provide on the cross sections can be easily translated to those asymmetries.

\subsection{Azimuthal Dependence}

In this work, we will fully exploit the azimuthal dependence of the DVCS and interference cross sections. One important observation about the azimuthal ($\phi$) dependence is that the cross sections can be written as finite-order polynomials of the trigonometric function $\cos(\phi)$ and $\sin(\phi)$, except for an overall azimuthal dependence in the denominator due to the BH propagators. This implies the existence of a maximal number of fitting parameters one can extract from the azimuthal dependence, requiring a more careful treatment of the azimuthal dependence. 

Based on this observation, we use a \textit{modified} cross section for which we simply re-arrange the azimuthal dependence in the denominator as,
\begin{equation}
    \sigma^{P_bP_t}(\phi) \to \frac{\mathcal P_1(\phi) \mathcal P_2(\phi)}{Q^4} \sigma^{P_bP_t}(\phi)\ ,
\end{equation}
where $\mathcal P_1(\phi)$ and $\mathcal P_2(\phi)$ are the BH propagators that cancel the azimuthal dependence in the denominator and an extra factor of $Q^{-4}$ is multiplie to preserve its dimension. We will discuss this operation with more details later in this section. Then the right-hand-side can be written in terms of finite-order polynomials of $\cos(\phi)$ and $\sin(\phi)$, which will permit a convenient access to the harmonic coefficients it possesses and allows a systematic approach to imposing direct experimental constraints to the CFFs found in the cross sections themselves.

A particular emphasis is given to cross sections in this study, with an additional look at the case of transversely-polarized target case done separately. The beam spin asymmetry will also be considered for CFF extraction, as these sorts of observables are also well-sought due to their often smaller experimental uncertainties.

\subsubsection{Logistical Overview}
As we just mentioned, since the \textit{modified} DVCS cross sections are finite-order polynomials of the cosine and sine trigonometric functions in $\phi$, we consider the following forms for the cross sections:
\begin{equation}
\label{eq:modxsection}
    \frac{\mathcal P_1(\phi) \mathcal P_2(\phi)}{Q^4} \sigma^{P_bP_t}_{\text{DVCS}+\mathcal{I}}= \sum_{n=0}^3\sigma_{\rm{mod},n}^{\rm{cos}}\cos(n\phi) +\sum_{n=1}^3\sigma_{\rm{mod},n}^{\rm{sin}}\sin(n\phi)\; ,
\end{equation}
where we suppress all the $(x_B,Q^2,t,y)$ dependence, as well as the $\phi_S$ dependence in the case of transverse polarized target, and we cut off the harmonics series to the third order which is the highest order that the twist-2 CFFs can get. Each $\sigma_{\rm{mod},n}^{\rm{cos}}$ or $\sigma_{\rm{mod},n}^{\rm{sin}}$ is the coefficient of the trigonometric function in the \textit{modified} cross sections that depends on the kinematics ($x_B,Q^2,t,y$) and the twist-2 CFFs $\mathcal F$, but NOT the azimuthal angle $\phi$. Therefore, given enough number and precision of measurements of different $\phi$ at the same kinematical point, each of those coefficients can be determined from their different azimuthal dependence, allowing us to put multiple constraints on the twist-2 CFFs if multiple coefficients are determined. 

Keeping this in mind, one would naturally think about exploiting the orthonormality property of the trigonometric functions (i.e. the foundation of Fourier analysis), as first discussed in \cite{Belitsky:2001ns}
 \begin{eqnarray}
 \frac{1}{\pi}\int_{-\pi}^\pi d\phi \cos(n\phi)\cos(m\phi)&=&\delta^{nm}\;,\label{orthocoscos}\\
 \frac{1}{\pi}\int_{-\pi}^\pi d\phi \sin(n\phi)\sin(m\phi)&=&\delta^{nm}\;,\label{orthosinsin}\\
 \frac{1}{\pi}\int_{-\pi}^\pi d\phi \sin(n\phi)\cos(m\phi)&=&0\; ,\label{orthosincos}
 \end{eqnarray}
 to extract each coefficient $\sigma_{\rm{mod},n}^{\rm{cos}}$ or $\sigma_{\rm{mod},n}^{\rm{sin}}$ from a physically measured cross section over $\phi$.
 Practically, we consider the following procedure:
 \begin{itemize}
 \item One can then imagine a DVCS experiment where the cross section is measured at many different $\phi$, keeping the kinematical point $(x_B,t,Q^2,y)$ fixed.  
 \item Using such data, one could subtract off the BH contribution and multiply the BH propagators $P_1(\phi) \mathcal P_2(\phi)/Q^4$ unambiguously and obtain data points of the  \textit{modified} DVCS cross section in eq. (\ref{eq:modxsection}) in terms of different $\phi$.
 \item One could then interpolate through the data points over $-\pi < \phi <\pi$, resulting in an experimental interpretation of the  \textit{modified} DVCS cross section. 
 \item Multiply this interpolating function by either $\sin(n\phi)$ or $\cos(n\phi)$ and integrate this product over $-\pi \leq \phi \leq \pi$. Due to eqs.(\ref{orthocoscos})-(\ref{orthosincos}), we get those harmonic coefficients $\sigma_{\rm{mod},n}^{\rm{cos}}$ or $\sigma_{\rm{mod},n}^{\rm{sin}}$ from the orthonormality property.
 \end{itemize}
  Then those measured (or extracted) $\sigma_{\rm{mod},n}^{\rm{cos}}$ and $\sigma_{\rm{mod},n}^{\rm{sin}}$ represent a system of equations which experimentally constrains our CFFs, as the relation between those coefficients and twist-2 CFFs are  determined in theory at the given kinematical point $(x_B,t,Q^2,y)$.
  
Such method indeed provides us a direct approach towards those coefficients $\sigma_{\rm{mod},n}^{\rm{cos}}$ and $\sigma_{\rm{mod},n}^{\rm{sin}}$, and also the harmonic coefficients of different order do not mix due to the orthonormality condition. However, this approach requires very precise measurements of the cross section with enough numbers of different $\phi$ in order to do the interpolation, which seems to be experimentally challenging.
Alternatively, to the same effect as the above procedure one could fit a harmonic function with free fitting coefficients, using the form in eq. (\ref{eq:modxsection}).
Once the parameters $\sigma_{\rm{mod},n}^{\rm{cos}}$ and $\sigma_{\rm{mod},n}^{\rm{sin}}$ are determined numerically, they may be equated to their expressions given in the Appendices, placing experimental constraints on the twist-2 CFFs.  We find this is much more convenient to do, allowing the use of the data point uncertainties, as the interpolation of the \textit{modified} DVCS cross section is best done by a fitting procedure anyway. In Section \ref{sec:3} we will arrange our cross sections in a convenient way in order to allow an ease of this extraction method.

\subsubsection{Harmonic Structure of Cross Sections}
The interference cross section contribution is inversely proportional to the Bethe-Heitler propagator factors $\mathcal{P}_1(\phi)\mathcal{P}_2(\phi)$, which will introduce additional complications to the overall azimuthal dependence of the total cross section.  These propagator factors are simply defined as
\begin{equation}\label{P1andP2}
    \mathcal{P}_1(\phi)=(k-q')^2\;,\;\;\; \mathcal{P}_2(\phi)=(k'+q')^2\;,
\end{equation}
 where $k(k')$ are the electron's initial(final) 4-momenta and $q'$ is the 4-momentum of the final state photon. To help us handle this term, we will express it into a harmonic series in itself as follows
\begin{equation}\label{P1P2harm}
    \frac{\mathcal{P}_1(\phi)\mathcal{P}_2(\phi)}{Q^4}=(BH)_0+(BH)_1\cos(\phi)+(BH)_2\cos(2\phi)\;,
\end{equation}
where
 \begin{eqnarray}
 (BH)_0&=&\frac{1}{y^2(1+\gamma^2)Q^4}\Bigg\{ 2M^2tx_B^2[y^2(x_B+1)-2y+2] +2M^4x_B^4y^2 \nonumber\\
 &~& +t^2[x_B(1-x_B)(y^2-6y+6)+y-1] +Q^2\Big[ 2M^2x_B^2(y^2-y+1)\nonumber\\
 &~& +t[(y-2)^2-x_B(y^2-6y+6)]\Big] +\frac{\gamma^2t}{2}\Big[2M^2x_B^2y^2 +t(x_B^2-x_B+1)y^2 \nonumber\\
 &~& -ty+t\Big] +\frac{\gamma^4t^2y^2}{8} + Q^4(y-1)\Bigg\} \;,\label{BH0}\\
 (BH)_1&=& -\frac{(y-2)(Q^2+tx_B)[Q^2+t(2x_B-1)]}{Q^3Mx_By^2(1+\gamma^2)^{3/2}}\sqrt{1-\frac{\bigg( \frac{\gamma^2(Q^2+t)}{2(tx_B+Q^2)} +1\bigg)}{1+\gamma^2}}\nonumber\\
 &~&\times \sqrt{1-y-\frac{\gamma^2y^2}{4}} \; ,\label{BH1}\\
 (BH)_2&=& -\frac{2\Big(y-1+\frac{\gamma^2y^2}{4} \Big)}{y^2(1+\gamma^2)Q^4}\Bigg[Q^2(M^2x_B^2-tx_B+t) +tx_B(2M^2x_B-tx_B+t) +\frac{\gamma^2t}{4} \Bigg]\;,\nonumber\\ \label{BH2}
 \end{eqnarray}
 which is in agreement with the BH propagators in both \cite{Kriesten:2019jep} and \cite{Belitsky:2010jw}. 
 
 

  We will define the Bethe-Heitler subtracted cross section by $\sigma_{\text{Tot-BH}}\equiv\sigma_{\text{DVCS}}+\sigma_\mathcal{I}$. We can then define the \textit{modified} cross section by multiplying the BH-subtracted one by the factor $\mathcal{P}_1\mathcal{P}_2$, which will reduce the azimuthal dependence to a simple, non-rational harmonic series for an easy extraction of its coefficients.
  
 \paragraph{UU Cross section}
 
 Then using the results of Appendix \ref{sec:A1} and \ref{sec:B1} together with Eq. (\ref{P1P2harm}) where the kinematical coefficients $h_n^{U}$ and $a_n^{I,\rm{U}},b_n^{I,\rm{U}}$ and $c_n^{I,\rm{U}}$ are defined, one can show that the reduced BH-subtracted $UU$ cross section is given by
 \begin{eqnarray}\label{sigUUlong}
 \frac{\mathcal{P}_1\mathcal{P}_2}{Q^4}\sigma^{UU}_{\text{Tot}-\text{BH}}&=&\frac{\Gamma}{Q^2}\frac{4}{Q^2}\sum_{i=0}^5(BH\otimes h^U)_i\mathcal{D}_1^{\text{DVCS}}\cos(i\phi)\nonumber \\
 &+&\frac{\Gamma}{Q^2t}\sum_{n=0}^3 \Bigg{\{}
 a_n^{I,\rm{U}} \mathcal{A}^U_{\text{Re}}+b_n^{I,\rm{U}} \mathcal{B}^U_{\text{Re}} + c_n^{I,\rm{U}}\mathcal{C}^U_{\text{Re}}
 \Bigg{\}}\cos(n\phi)\;,
 \end{eqnarray}
 where the new $BH\otimes h^U$ coefficients are given in Appendix \ref{App:E} and the CFF expressions $\mathcal{D}_1^{\text{DVCS}}$ and $\mathcal{\{A,B,C\}}^U_{\text{Re}}$ are defined in Appendix \ref{App:C}.

\paragraph{LU Cross section}
  
  This single spin cross section is unique as it only involves an interference contribution and is thus the only case which is purely linear in the CFFs. Consequently, it is among the simplest reduced cross sections we will see here. We have merely just to multiply out the BH propagators to get
  \begin{eqnarray}\label{sigLUlong}
  \frac{\mathcal{P}_1\mathcal{P}_2}{Q^4}\sigma^{LU}_{\text{Tot}-\text{BH}}&=& \frac{\Gamma}{Q^2t}(2h)\sum_{n=1}^3\Bigg{\{}
  a_n^{I,\rm{L}} \mathcal{A}^U_{\text{Im}}
  + b_n^{I,\rm{L}} \mathcal{B}^U_{\text{Im}} +
     c_n^{I,\rm{L}} \mathcal{C}^U_{\text{Im}}
  \Bigg{\}}\sin(n\phi)\; ,
  \end{eqnarray}
where $\mathcal{\{A,B,C\}}^U_{\text{Im}}$ are defined in Appendix \ref{App:C}.

\paragraph{UL Cross section}

We consider next the other single spin cross section with longitudinally polarized target.  This will be similar to the $LU$ case.  We will have
\begin{eqnarray}\label{sigULlong}
\frac{\mathcal{P}_1\mathcal{P}_2}{Q^4}\sigma^{UL}_{\text{Tot}-\text{BH}}&=&\frac{\Gamma}{Q^2t}(2\Lambda_l)\sum_{n=1}^3\sin(n\phi) \Bigg{\{} \widetilde{a}_n^{I,\rm{U}}\mathcal{A}^L_\text{Im} +\widetilde{b}_n^{I,\rm{U}} \mathcal{B}^L_\text{Im} + \widetilde{c}_n^{I,\rm{U}}\mathcal{C}^L_\text{Im}\Bigg{\}}\; ,
\end{eqnarray}
where $\mathcal{\{A,B,C\}}^L_{\text{Im}}$ are given in Appendix \ref{App:C}.

\paragraph{LL Cross section}

This polarization case is azimuthally analagous to the $UU$ case but with different coefficients and different CFF expressions.  We will have
\begin{eqnarray}\label{sigLLlong}
\frac{\mathcal{P}_1\mathcal{P}_2}{Q^4}\sigma^{LL}_{\text{Tot}-\text{BH}}&=&-8\frac{\Gamma}{Q^4}(2h)(2\Lambda_l)\sum_{j=0}^5 (BH\otimes h^{-,L})_j \mathcal{D}_2^\text{DVCS} \cos(j\phi)  \nonumber\\
    &~&+\frac{\Gamma}{Q^2t}(2h)(2\Lambda_l)\sum_{n=0}^3\cos(n\phi)\Bigg\{\tilde a_n^{I,\rm{L}}\mathcal{A}^L_\text{Re} +\tilde b_n^{I,\rm{L}} \mathcal{B}^L_\text{Re} +\tilde c_n^{I,\rm{L}}\mathcal{C}^L_\text{Re}\Bigg\}\; ,
\end{eqnarray}
where $\mathcal{D}_2^\text{DVCS}$ and $\mathcal{\{A,B,C\}}^L_{\text{Re}}$ are given in Appendix \ref{App:C}.

\subsubsection{Transverse Target Harmonics}

The presence of the factor $\cos(\phi_S-\phi)$ and $\sin(\phi_S-\phi)$ found in the $UT$ and $LT$ cross sections means that we already have a product of two trig functions in $\phi$.  If one then wishes to exploit the orthonormality properties of Eqs. (\ref{orthocoscos})-(\ref{orthosincos}), it is already possible to do so without multiplying the cross section by an additional sine or cosine function.  For example, the $UT,\rm{in}$ cross section has the form
\begin{eqnarray}\label{sigUTingen}
\sigma^{UT,\rm{in}}\sim\sum_m s^{UT,\rm{in}}_m\sin(m\phi)\cos(\phi_S-\phi)\;,
\end{eqnarray}
where $s^{UT,\rm{in}}_m$ is a kinematical constant multiplied by the twist-2 CFFs (due to both the DVCS and Interference parts in general).  Using trigonometric identities and integrating the cross section over the full azimuthal $\phi$-sphere then yields
\begin{equation}
    \frac{1}{\pi}\int_{-\pi}^{\pi}d\phi\sigma^{UT,\rm{in}}=s^{UT,\rm{in}}_1\sin(\phi_S)\;.
\end{equation}
Since the angle $\phi_S$ is event-dependent, we can view it as an independent angle which can be measured over through a sufficient spread of detected events.  An additional step then, would be to integrate over this angle to extract the coefficient $s_1^{UT,\rm{in}}$ via
\begin{eqnarray}
\frac{1}{2}\int_0^\pi d\phi_S \{ \sin{\phi_S,\cos\phi_S}\} = \{ 1,0 \}\;,\label{0pi_phiS}\\
\frac{1}{2}\int_{-\pi/2}^{\pi/2} d\phi_S \{ \sin{\phi_S,\cos\phi_S}\} = \{ 0,1 \}\;.\label{piby2_phiS}
\end{eqnarray}
%
%
%
%
%
In other words, for a transversely polarized target cross section, one does not need to multiply the cross section by a prescribed harmonic ($\cos(1\phi),\cos(2\phi),\it{etc.}$), but rather can directly integrate the cross section over the whole azimuthal sphere, which will only pull out the 1st harmonic coefficient.

We would like, however, to access all of the harmonic coefficients in cross sections such as (\ref{sigUTingen}).
To do this, we can simply regard a term like $\cos(\phi_S-\phi)$ as an independent term altogether. In other words, we can define the new azimuthal angle %
\begin{equation}
    \Delta\phi\equiv\phi_S-\phi\;,
\end{equation}
which would instead prescribe the appropriate integrations over $\Delta\phi$ to the effect of Eqs. (\ref{0pi_phiS}) \& (\ref{piby2_phiS}) and an additional integration over $\phi$ as in Eqs.(\ref{orthocoscos}-\ref{orthosincos}). We could indeed next proceed to construct $\mathcal{P}_1\mathcal{P}_2\sigma^{UT}_{\text{Tot-BH}}$ and $\mathcal{P}_1\mathcal{P}_2\sigma^{LT}_{\text{Tot-BH}}$ into a precise form as we did for $UU,UL,LU$ and $LL$.  However, we will wait to do this until Section \ref{sec:3}, where more practical information will reduce the number of harmonics worthy of consideration.

\subsubsection{Beam Spin Asymmetry}

We return now to the $A_{LR}$ asymmetry which was introduced early in this section.  Expressed in terms of harmonics, the asymmetry can be written as
%
%
\begin{eqnarray}\label{ALRfull}
A_{LR}\approx\frac{\sum_{n=1}^2 \Bigg[ \frac{a^{I,L}_n}{t}\mathcal{A}^U_{\text{Im}} + \frac{b^{I,L}_n}{t}\mathcal{B}^U_{\text{Im}} + \frac{c^{I,L}_n}{t}\mathcal{C}^U_{\text{Im}}\Bigg]\sin(n\phi)}{D_1 +D_2 + D_3},\nonumber\\
\end{eqnarray}
where we have pulled out a common factor of $\Gamma/(Q^2\mathcal{P}_1\mathcal{P}_2)$ from each of the present terms and we have removed the kinematically suppressed higher-twist terms in the numerator. Note that the asymmetry is overall parity odd in $\phi$ due to its numerator, but that being said, its denominator may still potentially give this observable a richer azimuthal structure. Each denominator term within is given by
\begin{eqnarray}
D_1&=&\frac{1}{Q^2t^2\Omega_0}\sum_{n=0}^2\Bigg[ a_n^{\text{BH}}\bigg(F_1^2-\frac{t}{4M^2}F_2^2 \bigg) + b_n^{\text{BH}}G_M^2 \Bigg]\cos(n\phi)\;,\label{den1}\\
D_2&=&\sum_{n=0}^2\frac{4}{Q^2}(BH\otimes h^U)_n\mathcal{D}_1^{\text{DVCS}}\cos(n\phi)\;,\label{den2}\\
D_3&=&\sum_{n=0}^2\Bigg[ \frac{a^{I,U}}{t}\mathcal{A}^U_{\text{Re}} + \frac{b^{I,U}}{t}\mathcal{B}^U_{\text{Re}} + \frac{c^{I,U}}{t}\mathcal{C}^U_{\text{Re}}\Bigg]\cos(n\phi)\;.\nonumber\label{den3}\\
\end{eqnarray}
where $\Omega_0$, $a_n^{\text{BH}}$ and $b_n^{\text{BH}}$ are all defined in Appendix \ref{App:D}. The terms $D_1$ and $D_3$ are in some regards the simplest, having just had their BH propagators removed while the term $D_2$ is only slightly more complicated, resulting in a \textit{compound} harmonic cosine series.


We have also removed the higher order harmonics $(n\geq3)$ which are highly suppressed and may be conveniently omitted.  It follows then, that $A_{LR}$ is a simple sine series divided by a simple cosine series.  Our next task with this, left to Section \ref{sec:3}, is to determine the least number of necessary harmonic coefficients needed to parametrize this asymmetry in a given kinematical range.

Another important point we make is that one is now forced to extract the harmonic coefficients in $A_{LR}$ differently than the way we have prescribed for the cross sections on account of the fact that we now have a ratio of series.  Once we have a fully reduced ratio of harmonic series for $A_{LR}$ using the above equations, we may fit an analogously parameterized function of $\phi$ directly to the data which represents $A_{LR}(\phi)$.  The parameterization would only require as many parameters as we have harmonic terms amongst the numerator and denominator.  We will perform this exercise in Section \ref{sec:4}.

\vspace{12pt}

\section{Reduced DVCS Observables}
\label{sec:3}

With all exact modified cross sections established in Section \ref{sec:2}, we now turn our focus towards further pragmatic considerations.  Although the harmonics-based method to observables is more than sufficient theoretically, the finite precision in DVCS cross section data may prove many of those observables unmeasurable. 

One important note is that those different harmonic coefficients are not equally relevant kinematically, as some of those harmonics functions (typically higher order ones) show up with coefficients that are kinematically suppressed by large ${Q^2}$. Motivated by this, we perform a complete power counting for all the DVCS observables listed in the last section. By dropping those harmonics that are not practically accessible, namely those come with kinematically suppressed coefficients which are intolerant to the contamination of other higher order effects we neglect in this work as well as the uncertainties of the experimental data themselves, we define our reduced DVCS observables which contains only those harmonics whose coefficients are non-zero at leading order.

We also note that in order to perform this systematical power counting, we have to perform kinematical twist expansion, for which we expand the coefficients with respect to large $Q^2$ order by order, such that we can find out the kinematical twist of each harmonics. Consequently, all our analysis are based on those kinematically twist expanded coefficients that have kinematically twist-4 accuracy. However, as discussed in ref. \cite{Guo:2021gru}, those higher order kinematical effects have quite comparable size for moderate $Q^2$, asking for a more precise treatment of those coefficients. Therefore, while those expanded coefficients are extremely helpful for our analysis in a general range of kinematics, we suggest using the all order kinematical coefficients in for instance ref. \cite{Belitsky:2010jw,Guo:2021gru} when extracting CFFs with experiment data, which involves only one specific kinematical point event by event.

In the following section, we will present our power counting results for ALL DVCS observables based on kinematical twist expansion and justified by explicit numerical evaluation. Those power counting results eventually define our reduced DVCS observables, allowing us to form a \textit{roadmap} which details which observables are ideal candidates for the extraction of each CFF.

\subsection{Kinematical Twist Considerations}

  Using the twist-4 expanded harmonic coefficients given in Appendices \ref{App:A} and \ref{App:B}, we have made Table \ref{HARobstable} to aid us in this study. The 3rd column of Table \ref{HARobstable} is informative in the following way.  As was shown in \cite{Braun:2011zr,Braun:2012hq,Braun:2014sta}, one can perform a more proper treatment of the twist expansion to the Compton Tensor, and thus to the DVCS and interference cross sections.  This will involve defining higher-order twist-2 CFFs, for example, accompanied by twist-3 kinematics.  Therefore, any cross section harmonic coefficient which starts at twist-3 kinematics or higher is contaminated with these corrections, which have been omitted from this analysis.  In fact, at the twist-3 level the genuine twist-3 CFFs also to enter the Compton tensor, and that is beyond the scope of this particular study.  It is believed, however, that the dynamical twist-3 terms are quite small compared to the dynamical twist-2 terms in DVCS observables.

\begin{table}[h]
    \centering
    \begin{tabular}{|Sc|Sc|Sc|}
    \hline
       Cross Section  & Coefficients & (Kin. Twist)$\times$Harmonics \\
       \hline
        \multirow{4}{*}{$\sigma^{UU}$ \& $\frac{\sigma^{UT,\rm{out}}}{\sin(\phi_S-\phi)}$} & \multirow{4}{*}{$h^U,A^{I,U},B^{I,U},C^{I,U}$} & $(\rm{tw}2+\rm{tw}3+\rm{tw}4)\cos(0\phi)$\\
         &  & $(\rm{tw}2+\rm{tw}3+\rm{tw}4)\cos(1\phi)$ \\
         &  & $(\rm{tw}3+\rm{tw}4)\cos(2\phi)$ \\
         &  & $(\rm{tw}4)\cos(3\phi)$ \\
        \hline
        \multirow{3}{*}{$\sigma^{UL}$ \& $\frac{\sigma^{UT,\rm{in}}}{\cos(\phi_S-\phi)}$ } &  \multirow{3}{*}{$\widetilde{A}^{I,U},\widetilde{B}^{I,U},\widetilde{C}^{I,U}$} & $(\rm{tw}2+\rm{tw}4)\sin(1\phi)$ \\
        ~ & ~ & $(\rm{tw}3)\sin(2\phi)$\\
        ~ & ~ & $(\rm{tw}4)\sin(3\phi)$\\
        \hline
        \multirow{2}{*}{$\sigma^{LU}$ \& $\frac{\sigma^{LT,\rm{out}}}{\sin(\phi_S-\phi)}$} & \multirow{2}{*}{$A^{I,L},B^{I,L},C^{I,L}$} & $(\rm{tw}2+\rm{tw}4)\sin(\phi)$\\
        ~ & ~ & $(\rm{tw}3)\sin(2\phi)$\\
        \hline
        \multirow{3}{*}{$\sigma^{LL}$ \& $\frac{\sigma^{LT,\rm{in}}}{\cos(\phi_S-\phi)}$} & \multirow{3}{*}{$h^{-,L},\widetilde{A}^{I,L},\widetilde{B}^{I,L},\widetilde{C}^{I,L}$} & $(\rm{tw}2+\rm{tw}3+\rm{tw}4)\cos(0\phi)$\\
        ~ & ~ & $(\rm{tw}2+\rm{tw}3+\rm{tw}4)\cos(1\phi)$\\
        ~ & ~ & $(\rm{tw}3)\cos(2\phi)$\\
        \hline
    \end{tabular}
    \caption{A list of the harmonic observables for each polarized DVCS cross section along with their hierarchy of kinematical twist contributions. Note that the out-of-plane  polarized cross section behaves the same as the unpolarized one in terms of kinematics except for an overall factor of $\sin(\phi_S-\phi)$. The same applies to the in-plane polarized cross-section and the longitudinally polarized one, except for an overall factor of $\cos(\phi_S-\phi)$ for the in plane polarized cross section.}
    \label{HARobstable}
\end{table}

However, at sufficiently large $Q^2$, we may begin to more justifiably view the cross section as being dominated by its twist-2 kinematics, and view the higher twist corrections as an uncertainty in the extraction of our 8 twist-2 CFFs.  Therefore, we shall only consider the harmonic coefficients which include leading twist kinematics, and regard all of the omitted higher twist terms as an additional source of uncertainty in our CFF extraction.  After all, the $Q^2$ dependence of the GPDs is not as interesting as the the $x_B$ and $t$-depenence of the GPDs, which defines them as non-perturbative distributions in the nucleon.

The information in Table \ref{HARobstable}, however, does not take into account the effect of the BH propagator interference effect we saw in Section \ref{sec:3}.  To make sense of the net twist content of our reduced cross sections, we now need to expand out the series sums and collect terms common to each harmonic level, then use the twist approximations from the Appendices.  To this end, we list the reduced cross sections, keeping only harmonic terms which include twist-2 kinematics.

\subsection{Reduced Cross Sections}

To begin, consider the general structure of a cross section with both a pure DVCS contribution and an interference contribution. We may express the reduced version of such a cross section generally by
\begin{eqnarray}\label{sigmaweights}
\frac{\mathcal{P}_1\mathcal{P}_2}{Q^4}\sigma_{\text{Tot-BH}}=\sigma_0\bigg[ \frac{(BH\otimes h^{\text{DVCS}})}{Q^2}\mathcal{D}(\mathcal{F}^2) + \frac{A^{\text{INT}}}{t}\mathcal{A}(\mathcal{F}) + \frac{B^{\text{INT}}}{t}\mathcal{B}(\mathcal{F}) + \frac{C^{\text{INT}}}{t}\mathcal{C}(\mathcal{F}) \bigg]\; , \nonumber\\
\end{eqnarray}
which can be applied to all polarization cases, and the cross section factor which is common to both DVCS and interference parts here is simply
\begin{equation}\label{sigma0}
    \sigma_0=\frac{\Gamma}{Q^2}\sim \frac{\text{nb}}{\text{GeV}^4}\;,
\end{equation}
and the remaining expression is completely unitless.  In fact, the CFF expressions themselves are also unitless.  Without any suspected numerical anomalies amongst $\mathcal{H,E,\widetilde{H},\widetilde{E}}$, we can think of the 4 unitless coefficients which accompany them as determining their relative weight (importance) towards the cross section. The supposition here is merely that no subset of the 8 twist-2 CFFs is orders of magnitude different than any of the remaining 8 ones. 

When we decompose the reduced cross sections into a harmonic series, each harmonic term in the series still preserves the general form of Eq.(\ref{sigmaweights}).  Their coefficients then, are in general functions of $(y,x_B,t,Q)$.  We can then look at the kinematical phase space expected to be spanned by certain experiments.  In particular, we choose here to look at the following JLab $12$~GeV kinematics \cite{CLAS:2015uuo}:
\begin{eqnarray}\label{kinematics}
&E_b=10.6~\text{GeV}\;,\nonumber\\
&Q^2=4.55~\text{GeV}^2\;,\nonumber\\
&0.1\leq x_B\leq 0.6\;,\nonumber\\
&|t_\text{min}|\leq -t\leq 0.8~\text{GeV}^2\;.
\end{eqnarray}

\begin{figure}[ht]
\centering

\begin{minipage}[b]{0.8\textwidth}
\centering
\includegraphics[width=0.8\textwidth]{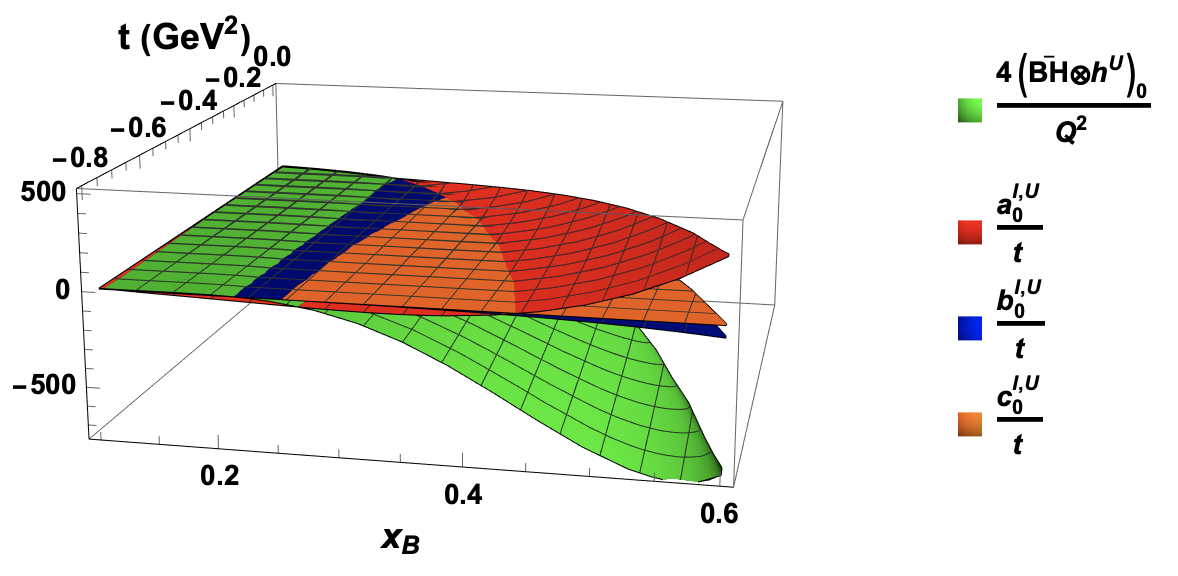}
\end{minipage}

\begin{minipage}[b]{0.8\textwidth}
\centering
\includegraphics[width=0.8\textwidth]{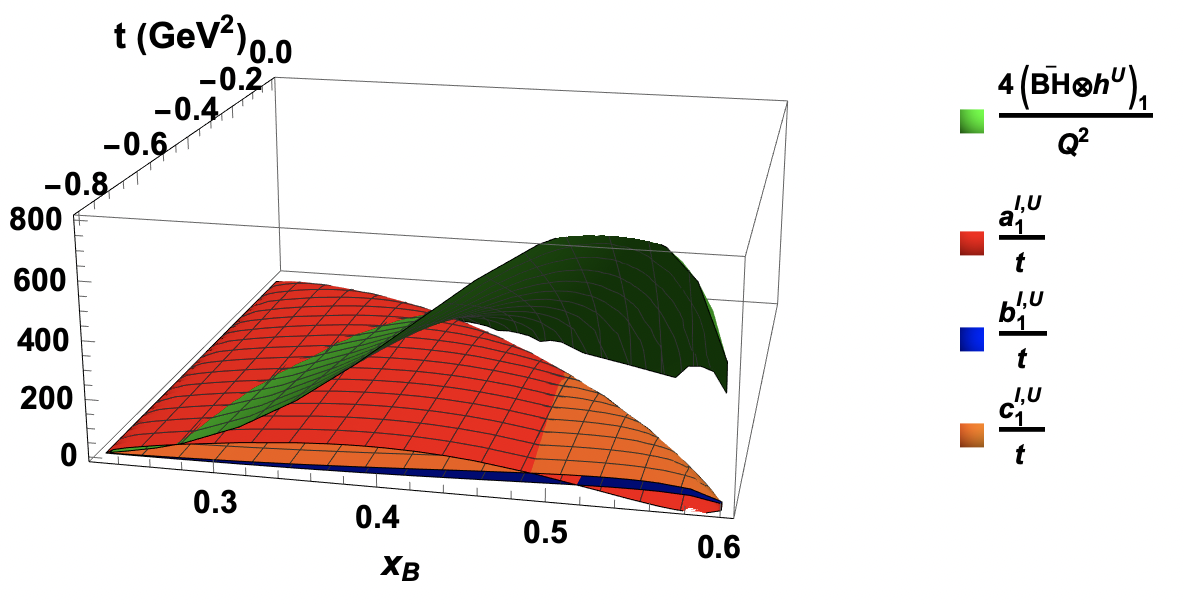}
\end{minipage}

\begin{minipage}[b]{0.8\textwidth}
\centering
\includegraphics[width=0.8\textwidth]{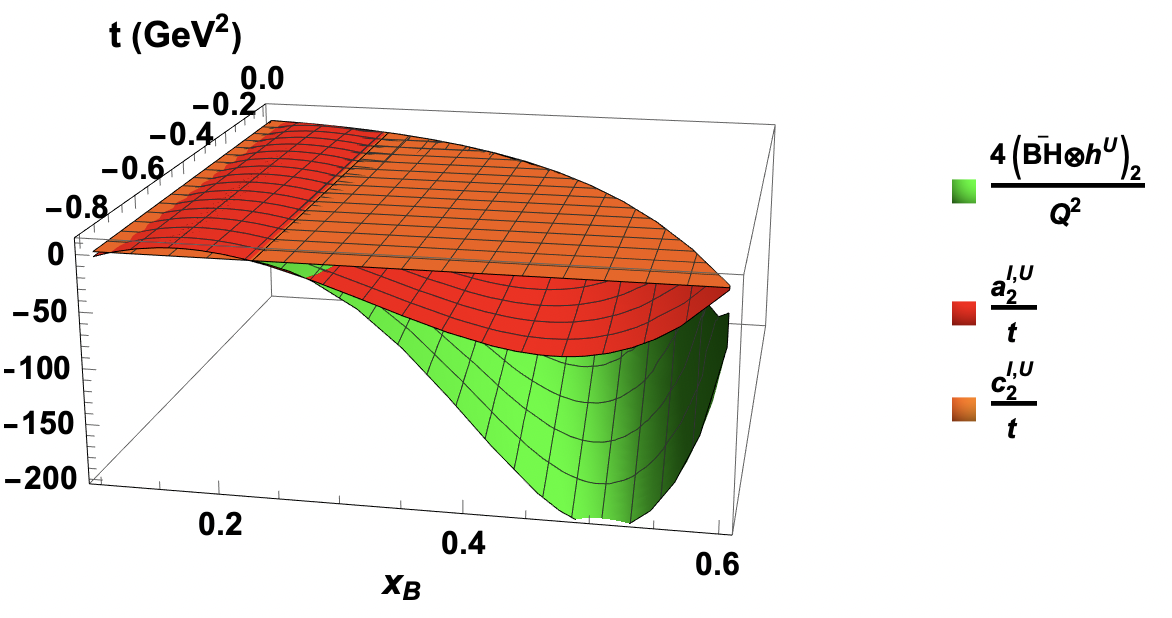}
\end{minipage}

\caption{\label{fig:UUsurfaceplots} The unitless kinematical coefficients present in the reduced UU cross section for harmonics $\cos(0\phi)$ (top), $\cos(1\phi)$ (middle) and $\cos(2\phi)$ (bottom).  They are plotted at the kinematics given in Eq.(\ref{kinematics})} 
\end{figure}

The choice to fix the beam energy and the $Q^2$ value is so that we may then think of the kinematic coefficients as a 2D surface in the $(x_B,t)$-plane. As one can see in Appendix \ref{App:A} \& \ref{App:B}, many of our kinematic coefficients depend on the factor $\kappa_t^{1/2}=\sqrt{(y-1)[t(x_B-1)-M^2x_B^2]}$.  Assuming $y\leq 1$, this places the following kinematical constraint between $x_B$ and $t$,
\begin{equation}\label{tmax}
    t\leq t_\text{min}=M^2\frac{x_B^2}{x_B-1}\;,
\end{equation}
which determines the minimum magnitude of $t$ given in Eq. (\ref{kinematics}). An important note here is that the general $Q^2$-dependent kinematical limits on $t$ are in fact somewhat different from the simpler Eq.(\ref{tmax}), and are given in \cite{Belitsky:2010jw} for example.  The reason we have a different (and less accurate) limit, is due to our twist-expanded coefficients, which will not have a noticeable effect on the numerical analysis. The unitless coefficients present in $\sigma^{UU}_{\cos(n\phi)}$ as an example, are plotted as surfaces in Figure \ref{fig:UUsurfaceplots}.

As one can see from Figure \ref{fig:UUsurfaceplots}, the overall magnitude of the $\cos(2\phi)$ coefficients is quite small compared to those of $\cos(1\phi)$ and $\cos(0\phi)$, and that in general the $a^{I,U}/t$ coefficient dominates the interference cross section over most of the $(t,x_B)$ plane.  It is reasonable then, to quantify this {\textit{hierarchy}} of contributions to each cross section in some meaningful and well-defined way.

We choose to compute the average value of each coefficient over its 2-dimensional plane in the same way one defines the average value of a function over its domain
\begin{equation}
    f_{\rm{avg}}=\frac{|\int_{dom}f(x_B,t)dx_B dt|}{\int_{dom}dx_B dt}\;,
\end{equation}
where $f(x_B,t)$ are the coefficients such as those plotted in Fig \ref{fig:UUsurfaceplots}, and $dom$ is the kinematically allowed domain of $(x_B,t)$ values.  The absolute value is purely conventional, as we are merely interested in the magnitude of each contribution to the cross section.  Since the harmonic factors $\cos(n\phi),\sin(n\phi)$ are of order unity and the CFF expressions involved for each given cross section polarization $(UU, LU, \text{etc.})$ are fixed, attributing Eq.(\ref{sigmaweights}) as {\textit{parts of the whole}} cross section is a meaningful way of determining the importance of each term.  Since we are trying to extract our CFFs, it is arguably imprudent to make any assumptions as to their numerical magnitude at this point.

We present the relative size of each coefficient as a percentage of its total reduced cross section in Table \ref{harmonicpercents}.

\begin{table}[h]
    \centering
    \begin{tabular}{|Sc|Sc|Sc|Sc|Sc|}
    \hline
       Harmonic  & \multicolumn{4}{Sc|}{Unitless contributions (\% whole cross section)} \\
       \hline
        $\sigma^{UU}_{\cos(0\phi)}$ &  $\frac{4(BH\otimes h^U)_0}{Q^2}\;(22\%)$& $\frac{a_0^{I,U}}{t}\;(14\%)$&$\frac{b_0^{I,U}}{t}\;(1\%)$&$\frac{c_0^{I,U}}{t}\;(7\%)$\\
        $\sigma^{UU}_{\cos(1\phi)}$ & $\frac{4(BH\otimes h^U)_1}{Q^2}\;(10\%)$&$\frac{a_1^{I,U}}{t}\;(35\%)$&$\frac{b_1^{I,U}}{t}\;(0.5\%)$&$\frac{c_1^{I,U}}{t}\;(4.5\%)$ \\
        $\sigma^{UU}_{\cos(2\phi)}$ & $\frac{4(BH\otimes h^U)_2}{Q^2}\;(1\%)$&$\frac{a_2^{I,U}}{t}\;(2.5\%)$&-&$\frac{c_2^{I,U}}{t}\;(0.5\%)$ \\
        \hline
        $\sigma^{LU}_{\sin(1\phi)}$ & -
        &$\frac{a_1^{I,L}}{t}\;(81\%)$
        &$\frac{b_1^{I,L}}{t}\;(0\%)$
        &$\frac{c_1^{I,L}}{t}\;(19\%)$ \\
        \hline
        $\sigma^{UL}_{\sin(1\phi)}$ &-
        & $\frac{\widetilde{a}_1^{I,U}}{t}\;(78\%)$
        &$\frac{\widetilde{b}_1^{I,U}}{t}\;(0\%)$
        &$\frac{\widetilde{c}_1^{I,U}}{t}\;(22\%)$ \\
        \hline
         $\sigma^{LL}_{\cos(0\phi)}$ & $\frac{8(BH\otimes h^{-,L})_0}{Q^2}\;(38\%)$&$\frac{\widetilde{a}_0^{I,L}}{t}\;(2.5\%)$&$\frac{\widetilde{b}_0^{I,L}}{t}\;(0.5\%)$&$\frac{\widetilde{c}_0^{I,L}}{t}\;(1.5\%)$ \\
        $\sigma^{LL}_{\cos(1\phi)}$ & $\frac{8(BH\otimes h^{-,L})_1}{Q^2}\;(36\%)$&$\frac{\widetilde{a}_1^{I,L}}{t}\;(11\%)$&$\frac{\widetilde{b}_1^{I,L}}{t}\;(0\%)$&$\frac{\widetilde{c}_1^{I,L}}{t}\;(3\%)$ \\
        $\sigma^{LL}_{\cos(2\phi)}$ & $\frac{8(BH\otimes h^{-,L})_2}{Q^2}\;(6\%)$&$\frac{\widetilde{a}_2^{I,L}}{t}\;(0.5\%)$&-&$\frac{\widetilde{c}_2^{I,L}}{t}\;(0\%)$ \\
        \hline
        $\sigma^{UT}_{\cos(0\phi)}$ & $\frac{4N(BH\otimes h^U)_0}{Q^2}\;(10\%)$&$\frac{2a_0^{I,U}}{Nt}\;(11.5\%)$&$\frac{2b_0^{I,U}}{Nt}\;(4\%)$&$\frac{2c_0^{I,U}}{Nt}\;(7\%)$ \\
       $\sigma^{UT}_{\cos(1\phi)}$ & $\frac{4N(BH\otimes h^U)_1}{Q^2}\;(12\%)$&$\frac{2a_1^{I,U}}{Nt}\;(30\%)$&$\frac{2b_1^{I,U}}{Nt}\;(3\%)$&$\frac{2c_1^{I,U}}{Nt}\;(5.5\%)$ \\
        $\sigma^{UT}_{\cos(2\phi)}$ & $\frac{4N(BH\otimes h^U)_2}{Q^2}\;(3\%)$&$\frac{2a_2^{I,U}}{Nt}\;(10\%)$&-&$\frac{2c_2^{I,U}}{Nt}\;(3\%)$ \\
        \hline
         $\sigma^{UT}_{\sin(1\phi)}$ & -
         &$\frac{2\widetilde{a}_1^{I,U}}{Nt}\;(78\%)$
         &$\frac{2\widetilde{b}_1^{I,U}}{Nt}\;(0\%)$
         &$\frac{2\widetilde{c}_1^{I,U}}{Nt}\;(22\%)$ \\
        \hline
      $\sigma^{LT}_{\cos(0\phi)}$ & $\frac{4N(BH\otimes h^{-,L})_0}{Q^2}\;(10\%)$&$\frac{2\widetilde{a}_0^{I,L}}{Nt}\;(7\%)$&$\frac{2\widetilde{b}_0^{I,L}}{Nt}\;(2\%)$&$\frac{2\widetilde{c}_0^{I,L}}{Nt}\;(6\%)$ \\
        $\sigma^{LT}_{\cos(1\phi)}$ & $\frac{4N(BH\otimes h^{-,L})_1}{Q^2}\;(9.5\%)$&$\frac{2\widetilde{a}_1^{I,L}}{Nt}\;(46\%)$&$\frac{2\widetilde{b}_1^{I,L}}{Nt}\;(1\%)$&$\frac{2\widetilde{c}_1^{I,L}}{Nt}\;(14\%)$ \\
        $\sigma^{LT}_{\cos(2\phi)}$ & $\frac{4N(BH\otimes h^{-,L})_2}{Q^2}\;(1.5\%)$&$\frac{2\widetilde{a}_2^{I,L}}{Nt}\;(2\%)$&-&$\frac{2\widetilde{c}_2^{I,L}}{Nt}\;(1\%)$ \\
        \hline
         $\sigma^{LT}_{\sin(1\phi)}$ & -
         &$\frac{2a_1^{I,L}}{Nt}\;(81\%)$
         &$\frac{2b_1^{I,L}}{Nt}\;(0\%)$
         &$\frac{2c_1^{I,L}}{Nt}\;(19\%)$ \\
        \hline
    \end{tabular}
    \caption{A table of numerical weights of each unit-less coefficient which contributes to all polarization harmonic cross sections. Weighting is performed over Eq.(\ref{kinematics}). Each term in the table represents the coefficient of a certain trigonometric function that can be explicitly written in terms of the kinematics. }
    \label{harmonicpercents}
\end{table}

The table suggests a few key points:

\begin{itemize}
    \item The 2nd order harmonics $\cos(2\phi)$ constitute a rather small percentage of their total cross sections, and thus it is likely not realistic to extract these terms from a measured cross section.
    \item For virtually any cross section harmonic, the $b^{\text{INT}}$ coefficient is usually at the sub-percent level, and can likely be omitted.  This general observation was reinforced in \cite{Guo:2021gru}. In fact, in some cases (e.g. Eq.(\ref{bILharmonics})) this coefficient is zero.
    \item For some harmonics the $c^{\text{INT}}$ coefficients are also quite small, for example in $\sigma^{LL}$.  This is because they are typically proportional to $\xi\; a^{\text{INT}}$ 
\end{itemize}

These numerical simplifications, which have only so far used the kinematical ranges of Eq.(\ref{kinematics}), are substantial, and it implies we effectively have less CFFs which practically enters Table \ref{CFFtable}.  Furthermore, one can consider the additional kinematical factors which reside in the expressions of Appendix \ref{App:C}. Using
\begin{eqnarray}
\xi_{\text{avg}}&\approx& 0.12\; , \\
\frac{t}{4M^2}\Bigg|_{\text{avg}}&\approx& 0.08\; ,
\end{eqnarray}
we can further approximate our reduced cross sections in Eqs.(\ref{sigUUapprox}-\ref{sigLTsinapprox}), revealing the highlighted dominant CFF terms shown in Table \ref{CFFtable}.  The reader may find a useful discussion of the following phenomenological consequences at the end of Section \ref{sec:3}.

With the above analysis, we may now propose expressions for reduced cross sections which do not include the suppressed higher-order harmonic coefficients.  

\paragraph{UU Cross Section:}

The reduced $UU$ cross section which includes leading twist kinematics is given by
\begin{eqnarray}\label{sigUUapprox}
\frac{\mathcal{P}_1\mathcal{P}_2}{Q^4}\sigma^{UU}_{\text{Tot-BH}}&\approx&\bigg[ \frac{4\Gamma}{Q^4}(BH\otimes h^U)_0\mathcal{D}_1^{\text{DVCS}} + \frac{\Gamma}{Q^2t}\big( a_0^{I,U}\mathcal{A}^U_{\rm{Re}} + c_0^{I,U}\mathcal{C}^U_{\rm{Re}}\big)\bigg]\cos(0\phi)\nonumber\\
&+&\bigg[ \frac{4\Gamma}{Q^4}(BH\otimes h^U)_1\mathcal{D}_1^{\text{DVCS}} + \frac{\Gamma}{Q^2t}\big( a_1^{I,U}\mathcal{A}^U_{\rm{Re}} + c_1^{I,U}\mathcal{C}^U_{\rm{Re}}\big)\bigg]\cos(1\phi)\; ,
\end{eqnarray}
where the unitless factor $\mathcal{P}_1\mathcal{P}_2/Q^4$  ensures that this reduced cross section still has the units of a real cross section.  Note that contrary to Table \ref{HARobstable}, the twist-2 kinematics actually {\it{leaks}} into the 2nd order harmonic coefficient due to the BH propagator interference effect.  However as shown in Section \ref{sec:5}, the $\cos(2\phi)$ coefficient still accounts for only $\sim 4\%$ of the $UU$ cross section and is hence removed.  The $b_n^{I,U}$ coefficients are also numerically negligible.

\paragraph{LU Cross Section:}

The reduced $LU$ cross section which includes leading twist kinematics is given by
\begin{eqnarray}\label{sigLUapprox}
\frac{\mathcal{P}_1\mathcal{P}_2}{Q^4}\sigma^{LU}_{\text{Tot-BH}}&\approx& \frac{\Gamma}{Q^2t}(2h)\bigg[ a_1^{I,L}\mathcal{A}^U_{\rm{Im}} +c_1^{I,L}\mathcal{C}^U_{\rm{Im}}\bigg] \sin(\phi)\; ,
\end{eqnarray}
where once again, the $b_n$-coefficients are negligible.

\paragraph{UL Cross Section:}

The reduced $UL$ cross section which includes leading twist kinematics is given by
\begin{eqnarray}\label{sigULapprox}
\frac{\mathcal{P}_1\mathcal{P}_2}{Q^4}\sigma^{UL}_{\text{Tot-BH}}&\approx& \frac{\Gamma}{Q^2t}(2\Lambda_l)\bigg[ \widetilde{a}_1^{I,U}\mathcal{A}^L_{\rm{Im}} +\widetilde{c}_1^{I,U}\mathcal{C}^L_{\rm{Im}}\bigg]\sin(\phi)\; ,
\end{eqnarray}
which is unsurprisingly quite similar to the approximated $LU$ cross section.

\paragraph{LL Cross Section:}

The reduced $LL$ cross section which includes leading twist kinematics is given by
\begin{eqnarray}\label{sigLLapprox}
\frac{\mathcal{P}_1\mathcal{P}_2}{Q^4}\sigma^{LL}_{\text{Tot-BH}}&\approx& (2h)(2\Lambda_l)\bigg\{ \bigg[ -\frac{8\Gamma}{Q^4}(BH\otimes h^{-,L})_0\mathcal{D}_2^{\text{DVCS}} + \frac{\Gamma}{Q^2t}\big( \widetilde{a}_0^{I,L}\mathcal{A}^L_{\rm{Re}} + \widetilde{c}_0^{I,L}\mathcal{C}^L_{\rm{Re}} \big)\bigg]\cos(0\phi)\nonumber\\
&~& \qquad\qquad +\bigg[ -\frac{8\Gamma}{Q^4}(BH\otimes h^{-,L})_1\mathcal{D}_2^{\text{DVCS}} + \frac{\Gamma}{Q^2t}\big( \widetilde{a}_1^{I,L}\mathcal{A}^L_{\rm{Re}} + \widetilde{c}_1^{I,L}\mathcal{C}^L_{\rm{Re}} )\bigg]\cos(1\phi)\nonumber\\
&~& \qquad\qquad  -\frac{8\Gamma}{Q^4}(BH\otimes h^{-,L})_2\mathcal{D}_2^{\text{DVCS}} \cos(2\phi)\bigg\}\; ,
\end{eqnarray}
  where again we have the twist-2 kinematics entering the 2nd order harmonic here through the DVCS term, just as it did for the UU case.  This time it is of the order of $6\%$ of the total $LL$ cross section, so we have kept it here.  Practically, were one to attempt to extract the $\cos(2\phi)$ coefficient from  $\sigma^{LL}$ data, it is somewhat doubtful this term will provide any useful constraint on $\mathcal{D}_2^\text{DVCS}$, as it would likely possess  a very large uncertainty.  Amongst the interference coefficients in Eq.(\ref{sigLLapprox}), it is $a^{I,L}_1$ which dominates, and will consequently account for a large portion of the \textit{peak-peak} amplitude in this cross section's $\phi$-variations.

\paragraph{UT Cross Section:}

We provide next the reduced $UT$ cross section which includes leading twist kinematics.  Due to its length, we will give each of its net harmonic coefficients separately.  We simply define these coefficients as follows,
\begin{equation}
    \frac{\mathcal{P}_1\mathcal{P}_2}{Q^4}\sigma^{UT}_{\text{Tot-BH}}=\sum_{n=0}^4\sigma^{UT}_{\cos(n\phi)}\cos(n\phi)\sin(\Delta\phi)+\sum_{n=1}^3\sigma^{UT}_{\sin(n\phi)}\sin(n\phi)\cos(\Delta\phi)\;.
\end{equation}
The cosine harmonic coefficients are given by
\begin{eqnarray}\label{sigUTcosapprox}
\sigma^{UT}_{\cos(n\phi)}&\approx&\frac{\Gamma}{Q^4}(2\Lambda_T)4N\mathcal{D}_3^{\text{DVCS}}(BH\otimes h^U)_n + \frac{\Gamma}{Q^2t}(2\Lambda_T)\frac{2}{N}(a_n^{I,U}\mathcal{A}_{\rm{Im}}^{\rm{out}}+c_n^{I,U}\mathcal{C}_{\rm{Im}}^{\rm{out}})\;,\;(n=0,1)\nonumber\\
&~&\\
\sigma^{UT}_{\cos(2\phi)}&\approx& \frac{\Gamma}{Q^2t}(2\Lambda_T)\frac{2}{N}a_2^{I,U}\mathcal{A}_{\rm{Im}}^{\rm{out}}\; ,
\end{eqnarray}
with the higher order harmonics starting at sub-leading twist or higher and $\mathcal{D}_3^\text{DVCS}$ and $\mathcal{\{A,C\}}^\text{out}_\text{Im}$ defined in Appendix \ref{App:C}.  Once again, the $\cos(2\phi)$ term would require very precise data in order to place a useful constraint on the twist-2 CFFs.  Meanwhile the sine harmonic coefficients are given by
\begin{eqnarray}\label{sigUTsinapprox}
\sigma^{UT}_{\sin(1\phi)}\approx -\frac{\Gamma}{Q^2t}(2\Lambda_T)\frac{2}{N}(\widetilde{a}_1^{I,U}\mathcal{A}_{\rm{Im}}^{\rm{in}}+\widetilde{c}_1^{I,U}\mathcal{C}_{\rm{Im}}^{\rm{in}})\; ,
\end{eqnarray}
where $\mathcal{\{A,C\}}^\text{in}_\text{Im}$ are given in Appendix \ref{App:C} and the higher order sine harmonics do not contain any leading twist kinematics -- and are thus not considered here.

\paragraph{LT Cross Section:}

The reduced $LT$ cross section will be similar to the $UT$ case, and we again define the net harmonic coefficients of the $LT$ cross section via
\begin{equation}
     \frac{\mathcal{P}_1\mathcal{P}_2}{Q^4}\sigma^{LT}_{\text{Tot-BH}}=\sum_{n=0}^4\sigma^{LT}_{\cos(n\phi)}\cos(n\phi)\cos(\Delta\phi)+\sum_{n=1}^3\sigma^{LT}_{\sin(n\phi)}\sin(n\phi)\sin(\Delta\phi)\;,
\end{equation}
in which the cosine coefficients are given by
\begin{eqnarray}\label{sigLTcosapprox}
\sigma^{LT}_{\cos(n\phi)}&\approx&-\frac{\Gamma}{Q^4}(2h)(2\Lambda_T)(4N)(BH\otimes h^{-,L})_n\mathcal{D}_4^{\text{DVCS}} + \frac{\Gamma}{Q^2t}(2h)(2\Lambda_T)\frac{2}{N}\big( \widetilde{a}_n^{I,L}\mathcal{A}_{\rm{Re}}^{\rm{in}} +\widetilde{c}_n^{I,L}\mathcal{C}_{\rm{Re}}^{\rm{in}}\big)\;, \; (n=0,1)\nonumber\\
\end{eqnarray}
while the sine coefficients are
\begin{equation}\label{sigLTsinapprox}
    \sigma^{LT}_{\sin(1\phi)}\approx -\frac{\Gamma}{Q^2t}(2h)(2\Lambda_T)\frac{2}{N}\big( a_1^{I,L}\mathcal{A}_{\rm{Re}}^{\rm{out}} + c_1^{I,L}\mathcal{C}_{\rm{Re}}^{\rm{out}}\big)\;,
\end{equation}
where $\mathcal{D}_4^{\text{DVCS}}$ and $\{\mathcal{A,C}\}^{\{\rm{in,out}\}}_{\rm{Re}}$ are defined in Appendix \ref{App:C}.

\subsection{Compton Form Factor Map}

With the results of the previous subsection, we can summarize a map of where each of the eight twist-2 CFFs enters our observables into Table \ref{CFFtable}. As we can see from the table, all 8 CFFs are spanned by the 8 cross sections, none of which is particularly sensitive to $\widetilde{\mathcal{E}}$.  Charge asymmetries may allow one to further decouple the pure DVCS from the interference part, allowing more explicit access to the real and imaginary parts of our CFFs, but we are not considering these observables in this study.  The dominant terms highlighted in Table \ref{CFFtable} have phenomenological consequences which we shall now discuss.

\begin{table}[h]
    \centering
    \begin{tabular}{|Sc|Sc|Sc|Sc|Sc|Sc|Sc|Sc|Sc|Sc|}
    \hline
       Harmonic & Expressions & $\text{Re}\mathcal{H}$ & $\text{Im}\mathcal{H}$ & $\text{Re}\mathcal{E}$ & $\text{Im}\mathcal{E}$ & $\text{Re}\widetilde{\mathcal{H}}$ & $\text{Im}\widetilde{\mathcal{H}}$ & $\text{Re}\widetilde{\mathcal{E}}$ & $\text{Im}\widetilde{\mathcal{E}}$ \\
       \hline
    $\sigma^{UU}_{\cos(n\phi)}$ & $\mathcal{D}_1^{\text{DVCS}},\mathcal{A}^U_{\rm{Re}},\mathcal{B}^U_{\rm{Re}},\mathcal{C}^U_{\rm{Re}}$ & {\color{red}\ding{52}} & {\color{red}\ding{52}} & \ding{51} & \ding{51} & {\color{red}\ding{52}} & {\color{red}\ding{52}} & \ding{51} & \ding{51}    \\
    \hline
     $\sigma^{LU}_{\sin(1\phi)}$  & $\mathcal{A}^U_{\rm{Im}},\mathcal{B}^U_{\rm{Im}},\mathcal{C}^U_{\rm{Im}}$ & - & {\color{red}\ding{52}} & - & \ding{51} & - & {\color{red}\ding{52}} & - & - \\
     \hline
       $\sigma^{UL}_{\sin(1\phi)}$ & $\mathcal{A}^L_{\rm{Im}},\mathcal{B}^L_{\rm{Im}},\mathcal{C}^L_{\rm{Im}}$ & - & {\color{red}\ding{52}} & - & \ding{51} & - & {\color{red}\ding{52}} & - & \ding{51} \\
       \hline
       $\sigma^{LL}_{\cos(n\phi)}$ & $\mathcal{D}_2^{\text{DVCS}},\mathcal{A}^L_{\rm{Re}},\mathcal{B}^L_{\rm{Re}},\mathcal{C}^L_{\rm{Re}}$ & {\color{red}\ding{52}} & {\color{red}\ding{52}} & \ding{51} & \ding{51} & {\color{red}\ding{52}} & {\color{red}\ding{52}} & \ding{51} & \ding{51} \\
       \hline
       $\sigma^{UT}_{\cos(n\phi)}$ & $\mathcal{D}_3^{\text{DVCS}},\mathcal{A}^{\rm{out}}_{\rm{Im}},\mathcal{B}^{\rm{out}}_{\rm{Im}},\mathcal{C}^{\rm{out}}_{\rm{Im}}$ & \ding{51} & {\color{red}\ding{52}} & \ding{51} & {\color{red}\ding{52}} & \ding{51} & \ding{51} & \ding{51} & \ding{51} \\
       \hline
       $\sigma^{UT}_{\sin(1\phi)}$ & $\mathcal{A}^{\rm{in}}_{\rm{Im}},\mathcal{B}^{\rm{in}}_{\rm{Im}},\mathcal{C}^{\rm{in}}_{\rm{Im}}$ & - & {\color{red}\ding{52}} & - & {\color{red}\ding{52}} & - & {\color{red}\ding{52}} & - & \ding{51} \\
       \hline
       $\sigma^{LT}_{\cos(n\phi)}$ & $\mathcal{D}_4^{\text{DVCS}},\mathcal{A}^{\rm{in}}_{\rm{Re}},\mathcal{B}^{\rm{in}}_{\rm{Re}},\mathcal{C}^{\rm{in}}_{\rm{Re}}$ & {\color{red}\ding{52}} & \ding{51} & {\color{red}\ding{52}} & \ding{51} & {\color{red}\ding{52}} & \ding{51} & \ding{51} & \ding{51} \\
       \hline
        $\sigma^{LT}_{\sin(1\phi)}$ & $\mathcal{A}^{\rm{out}}_{\rm{Re}},\mathcal{B}^{\rm{out}}_{\rm{Re}},\mathcal{C}^{\rm{out}}_{\rm{Re}}$ & {\color{red}\ding{52}} & - & {\color{red}\ding{52}} & - & {\color{red}\ding{52}} & - & \ding{51} & - \\
        \hline
    \end{tabular}
    \caption{A list of the CFF content of each of the leading twist harmonic cross sections. Cells containing larger, red check marks indicate a stronger dependence on that CFF, while the smaller black check marks indicate subdominant terms in the cross section.}
    \label{CFFtable}
\end{table}

The dominant CFF terms found in each polarization cross section are as follows
\begin{eqnarray}\label{CFFmapping}
\sigma^{UU}_{\cos(n\phi)}&\sim& \text{Re}\mathcal{H},\text{Im}\mathcal{H},\text{Re}\widetilde{\mathcal{H}},\text{Im}\widetilde{\mathcal{H}} \qquad\qquad\; (n=0,1)\nonumber\\
\sigma^{LU}_{\sin(1\phi)}&\sim& \text{Im}\mathcal{H} ,  \text{Im}\widetilde{\mathcal{H}} \nonumber\\
\sigma^{UL}_{\sin(1\phi)}&\sim& \text{Im}\mathcal{H} , \text{Im}\widetilde{\mathcal{H}} \nonumber\\
\sigma^{LL}_{\cos(n\phi)}&\sim& \text{Re}\mathcal{H},\text{Im}\mathcal{H},\text{Re}\widetilde{\mathcal{H}},\text{Im}\widetilde{\mathcal{H}} \qquad\qquad\; (n=0,1)\\
\sigma^{UT}_{\cos(n\phi)}&\sim& \text{Im}\mathcal{H} ,  \text{Im}\mathcal{E} \qquad\qquad\qquad\qquad\;\; (n=0,1)\nonumber\\
\sigma^{UT}_{\sin(1\phi)}&\sim& \text{Im}\widetilde{\mathcal{H}} , \text{Im}\mathcal{E}, \text{Im}\mathcal{H} \nonumber\\
\sigma^{LT}_{\cos(n\phi)}&\sim& \text{Re}\widetilde{\mathcal{H}},  \text{Re}\mathcal{E}, \text{Re}\mathcal{H} \qquad\qquad\qquad\;\; (n=0,1)\nonumber\\
\sigma^{LT}_{\sin(1\phi)}&\sim& \text{Re}\mathcal{H} , \text{Re}\mathcal{E}, \text{Re}\widetilde{\mathcal{H}}\;,\nonumber
\end{eqnarray}
which suggests a natural observable set for extracting the CFFs $\mathcal{H}$ and $\widetilde{\mathcal{H}}$:
\begin{equation}
    \{\sigma^{LU},\sigma^{UL},\sigma^{UU},\sigma^{LL}\}\rightarrow  \mathcal{H},\widetilde{\mathcal{H}}\;.
\end{equation}
Once this has been done, the CFF $\mathcal{E}$ has several possible measurements for its extraction, namely:
\begin{eqnarray}
    \sigma^{UT}&\rightarrow& \text{Im}\mathcal{E}\; ,\\
    \sigma^{LT}&\rightarrow& \text{Re}\mathcal{E}\; .
\end{eqnarray}
In a more recent study in \cite{Kriesten:2020apm}, a generalized Rosenbluth  separation technique on $\sigma^{UU}$ and $\sigma^{LU}$ has been proposed for the extraction of both $\mathcal{H}$ and $\mathcal{E}$.  This technique requires, however: 1) the approximation that the pure DVCS cross section is $\phi$-independent, 2) the interference cross section contains no zeroth-order harmonic terms and 3) that the $C^{I,U}$ terms are negligible with respect to both the $A^{I,U}$ and $B^{I,U}$ terms.  The first of these requirements comes at a cost in accuracy which is difficult to quantify, while we do not observe the latter 2 conditions. 

Unfortunately, the GPD $\widetilde{\mathcal{E}}$ is always suppressed by kinematical factors (for example terms proportional to $\xi$) in all of the cross sections, and may therefore prove more challenging to extract. Nevertheless, $\text{Im}\widetilde{\mathcal{E}}$ can be accessed through $\sigma^{UL},\sigma^{UT}_{\sin\phi}$, while $\text{Re}\widetilde{\mathcal{E}}$ through $\sigma^{UU},\sigma^{LL}$ as well as some of the other transversely-polarized target harmonics (see Table \ref{CFFtable}). We shall numerically study the importance of the various polarization cross sections towards CFF extraction in Section \ref{sec:4}.

\subsection{Reduced Asymmetries}

We can repeat the same technique of averaging our coefficients over JLab 12 GeV kinematical phase space to those terms in the beam spin asymmetry $A_{LR}$.  Doing so reveals that the higher order harmonics drop out due to kinematical suppression, leaving us with the formula,
\begin{eqnarray}\label{ALRphen}
A_{LR}\approx\frac{\alpha_1\sin\phi + \alpha_2\sin(2\phi)}{1+\beta_1\cos\phi+\beta_2\cos(2\phi)}\; ,
\end{eqnarray}
where the coefficients of the numerator are approximately given by
\begin{eqnarray}
\alpha_1&\approx& \frac{1}{\beta_0}\bigg[ \frac{a^{I,L}}{t}\mathcal{A}^U_{\text{Im}} +  \frac{c^{I,L}}{t}\mathcal{C}^U_{\text{Im}}\bigg]\;, \\
\alpha_2&\approx& \frac{1}{\beta_0}\bigg[  \frac{a^{I,L}}{t}\mathcal{A}^U_{\text{Im}} + \frac{c^{I,L}}{t}\mathcal{C}^U_{\text{Im}}\bigg]\;,
\end{eqnarray}
and
\begin{equation}\label{beta0}
    \beta_0= \frac{a_0^{\text{BH}}}{Q^2t^2\Omega_0}\bigg( F_1^2-\frac{t}{4M^2}F_2^2\bigg)+\frac{b_0^{\text{BH}}}{Q^2t^2\Omega_0}G_M^2 +\frac{4}{Q^2}(\overline{BH}\otimes h^U)_0\mathcal{D}_1^\text{DVCS} + \frac{a_0^{I,U}}{t}\mathcal{A}^U_{\text{Re}}+\frac{c_0^{I,U}}{t}\mathcal{C}^U_{\text{Re}}\;.
\end{equation}
While the coefficients of the denominator $(\beta_1,\beta_2)$ are identical to Eq. (\ref{beta0}) but with the next order $(1,2)$ coefficients inserted respectively, and are also divided by the factor $\beta_0$.
%
%
Indeed the BH cross section terms dominate within $A_{LR}$, however the interference terms are found to contribute non-negligibly to the denominator, and thus we have kept such terms (proportional to $\mathcal{A}^U_{\text{Re}}$) in the $\beta$ coefficients.  The DVCS terms must also be kept in.

A measurement of $A_{LR}$ is a way to extract the imaginary part of the CFFs $\mathcal{H,E,\widetilde{H}}$.  Beyond that, sufficiently precise data for this asymmetry may allow the extraction of the real parts of $\mathcal{H,E}$ through its present $\mathcal{\{A,C\}}^U_{\text{Re}}$ terms, although this remains to be seen.  Therefore, at the very least we can expect
\begin{equation}
    A_{LR}\rightarrow \text{Im}\{ \mathcal{H,E,\widetilde{H}} \}\;.
\end{equation}
An important note about the asymmetry, is that some of the coefficients present here do not involve leading twist terms, meaning they are missing additional terms that are potentially just as large.  The reason we cannot remove these sub-leading twist terms and only use leading twist terms is because we do not have the luxury of explicitly (or implicitly) integrating out one specific harmonic coefficient from the order-unity $A_{LR}$ as we did for the cross sections in the previous section.  It follows then, that a more extensive twist-3 analysis of $A_{LR}$ is worthy of future consideration.

\section{Numerical Analysis}
\label{sec:4}

We now undertake the task of applying our CFF extraction method(s) on both existing JLab $12~$GeV cross section data as well as model-computed pseudo-data.  We do not yet possess a sufficient set of observables to extract all 8 twist-2 CFFs, and so we attempt to answer first the question: how well can one detect or isolate each harmonic coefficient with real data?  Following this, we shall study the fitting of harmonics to pseudo-data, covering all the possible polarization cross sections.  Given those numerical constraints on the CFFs we will finish by extracting all eight with nonlinear least-squares methods.

\subsection{Unpolarized Cross Section Data Test}

\begin{figure}[ht]
\centering
\includegraphics[width=0.70\textwidth]{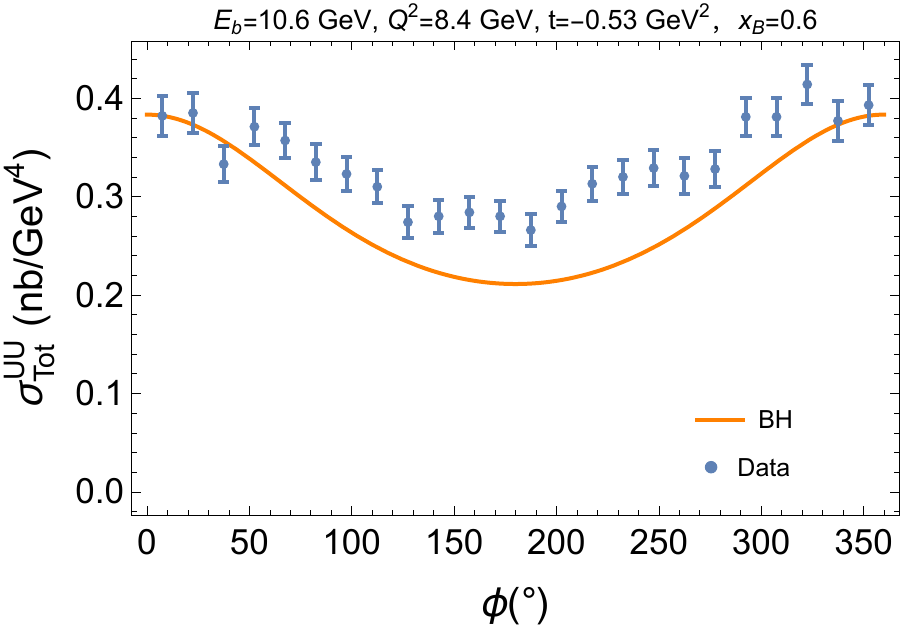}
\caption{\label{sigUUdata} Data points for $\sigma^{UU}$ from one particular bin of data from \cite{GeorgesPhD}.  Shown also is the calculated Bethe-Heitler cross section (using the results of Appendix \ref{App:D}) at the same kinematics.} 
\end{figure}

Next, let's perform the exercise of fitting harmonic coefficients to real DVCS cross section data.  We will choose a set of data for the unpolarized cross section $\sigma^{UU}$ provided in \cite{GeorgesPhD} at the kinematical point: $E_b=10.6~\text{GeV}$, $Q^2=8.4~\text{GeV}^2$, $x_B=0.6$, $t=0.76~\text{GeV}^2$ taken over $0<\phi<2\pi$ shown in Figure \ref{sigUUdata}.  From here we must subtract from the data the calculated BH contribution.  Then one has the choice of either trying to fit a harmonic parameterization directly to $\sigma^{UU}_{\text{Tot-BH}}$ or multiply out the BH propagators and fit a harmonic parameterization to $\frac{\mathcal{P}_1\mathcal{P}_2}{Q^4}\sigma^{UU}_{\text{Tot-BH}}$.  The result of either choice is shown in Figure 3.

Each cross section contains the following (fully reduced) azimuthal dependence,
\begin{eqnarray}
\sigma^{UU}_{\text{Tot-BH}}&=&\frac{c_0+c_1\cos\phi}{(BH)_0+(BH)_1\cos\phi+(BH)_2\cos(2\phi)}\; , \label{UUtotmbhparam}\\
\frac{\mathcal{P}_1\mathcal{P}_2}{Q^4}\sigma^{UU}_{\text{Tot-BH}}&=& \widetilde{c}_0+\widetilde{c}_1\cos\phi \;,\label{UUmodparam}
\end{eqnarray}
where the unitless BH coefficients were given explicitly in Section 3 and are exactly calculable. 
On one hand, we can simply think of the extra $\mathcal{P}_1\mathcal{P}_2$ factor as \textit{extra baggage} that is simply dealt with.  On the other hand, this factor possesses the observed qualitative effect of broadening the full width half maximum (FWHM) of the central peak in the BH-subtracted cross section, which better fits the overall shape of the data in $\phi$ (e.g. the curve in Figure 3 left is visibly broader than that in Figure 3 right).

\begin{figure}
\centering
\begin{minipage}{.5\textwidth}
  \centering
  \includegraphics[width=.9\linewidth]{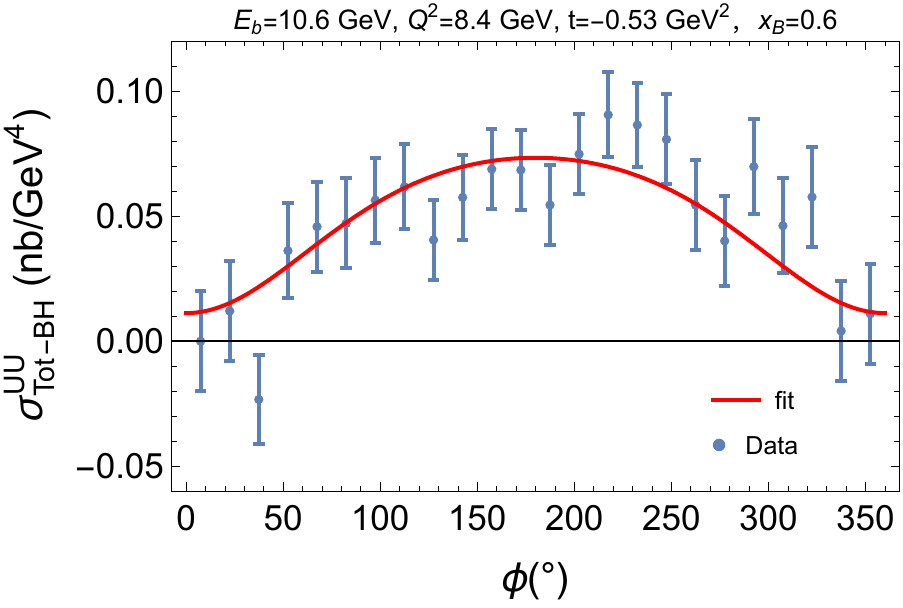}
  \label{fig:test1}
\end{minipage}%
\begin{minipage}{.5\textwidth}
  \centering
  \includegraphics[width=.9\linewidth]{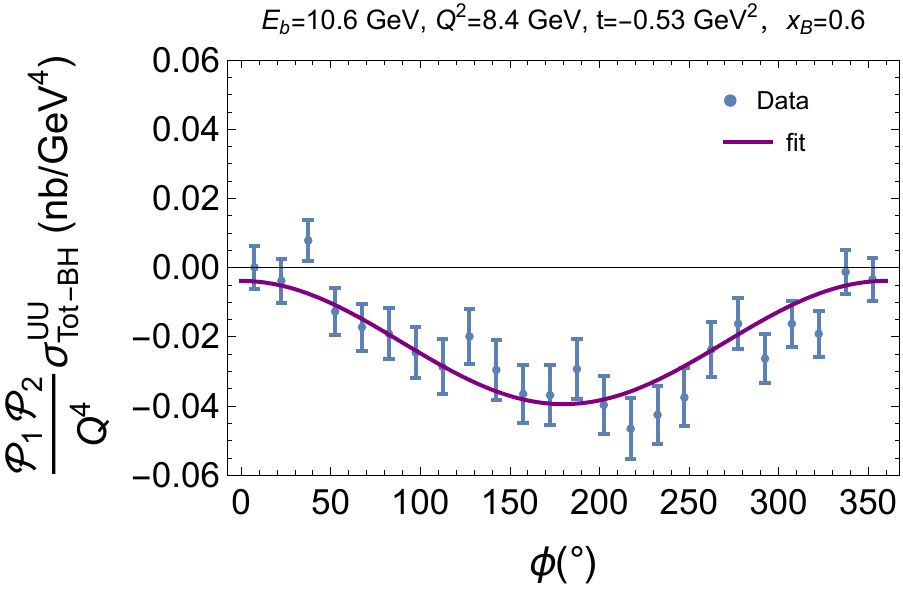}
  \label{fig:test2}
\end{minipage}
\caption{\label{fig:UUfittedplots} Comparison of fitting a harmonic parameterization to both the BH-subtracted UU cross section (left) and the reduced UU cross section (right) which was discussed in Section \ref{sec:3}.} 
\end{figure}

Fortunately, the simple harmonic form of the reduced unpolarized cross section may in fact be reduced to a linear fitting exercise.  Defining the new linear variable
\begin{equation}
    x_p=\begin{cases}\cos\phi ,\;\; \pi< \phi < 2\pi \\ -\cos\phi , 0 < \phi < \pi \end{cases}
\end{equation}
allows us to perform a traditional least-squares linear regression fit of both halves of the data,
\begin{eqnarray}
\frac{\mathcal{P}_1\mathcal{P}_2}{Q^4}\sigma^{UU}_{\text{Tot-BH}}\Bigg|_{\phi\geq \pi}=\widetilde{c}_0+\widetilde{c}_1 x_p\;, \label{rightxp} \\
\frac{\mathcal{P}_1\mathcal{P}_2}{Q^4}\sigma^{UU}_{\text{Tot-BH}}\Bigg|_{\phi \leq \pi}=\widetilde{c}_0-\widetilde{c}_1 x_p\;.\label{leftxp}
\end{eqnarray}

The results of these two fits is shown in Figure 4.  One expects a symmetry of the data about $\phi=\pi$, as well as the consistency of the slopes and intercepts between the two fitted lines.  Geometrically, the zeroth order harmonic coefficient is the y-intercept and the average slope (absolute value) represents the 1st order harmonic coefficient.  The uncertainties of the fitted parameters is now $\sim10\%$.

\begin{figure}[ht]
\centering
\includegraphics[width=0.75\textwidth]{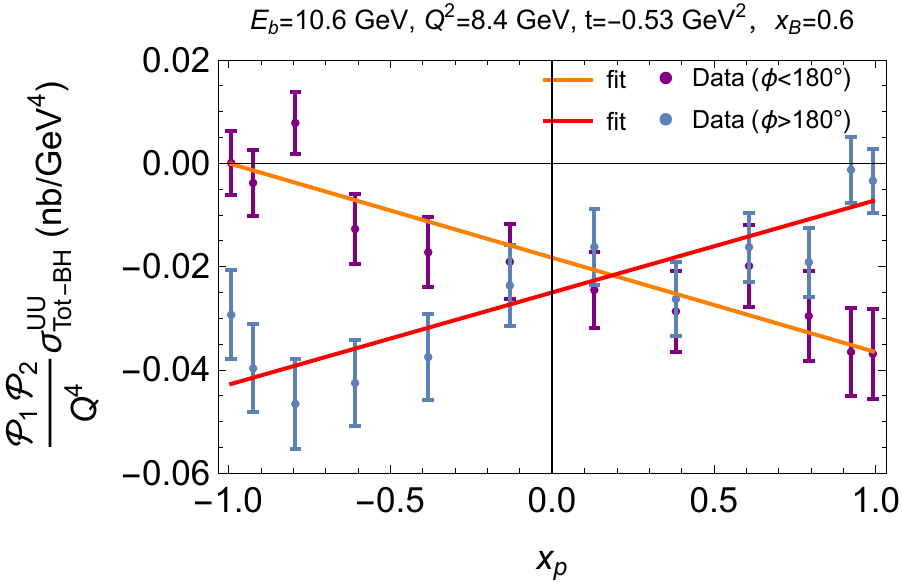}
\caption{\label{sigUUlinearizedplot} The results of fitting Eqs. (\ref{rightxp}),(\ref{leftxp}) to the first bin of data for kinematical setting 603 from \cite{GeorgesPhD}} 
\end{figure}

The fact that we can successfully fit both the zeroth and first order harmonic coefficients to this particular set of $UU$ cross section data to within $\sim 10\%$ is encouraging. It is indeed fortunate that $\sigma^{UU}$ data can be reduced to a simple 1st order harmonic function given in Eq.(\ref{UUmodparam}), and in fact this form is apparent upon inspection of the data (see Figure 3 right).  This allows a linearization procedure, which gives us a very clear geometrical extraction of two experimentally-determined constants which constrains our twist-2 CFFs.  For the case of a singly-polarized cross section, the same procedure can be employed, resulting in a linear relationship with a zero y-intercept and $x_p\rightarrow \sin\phi$ (for example see Eq.(\ref{sigLUapprox})).


\subsection{Beam Spin Asymmetry Data Test}

\begin{figure}[ht]
\centering
\includegraphics[width=0.6\textwidth]{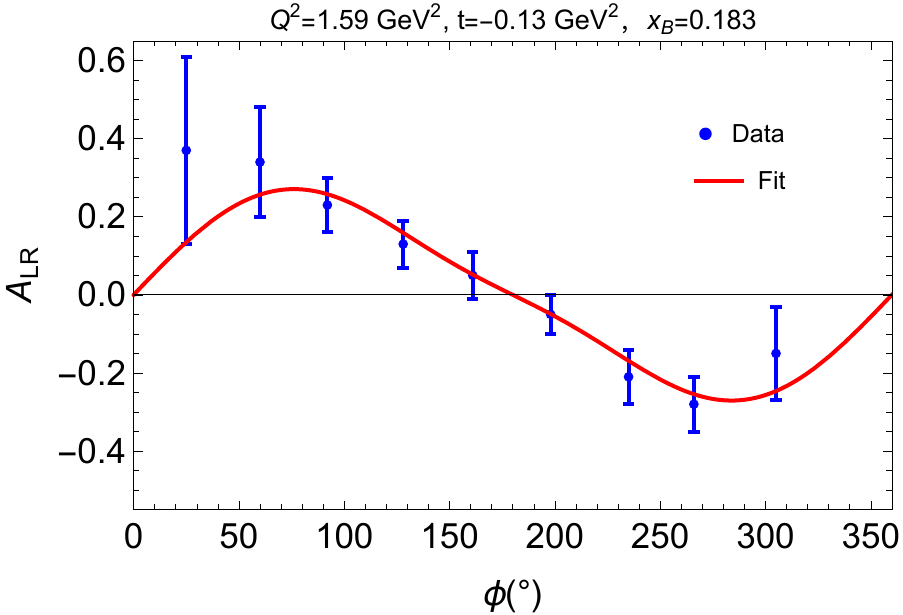}
\caption{\label{ALRfit} The result of fitting Eq.(\ref{eq:ALRfinal}) to one particular bin of data from \cite{CLAS:2015bqi}} 
\end{figure}

We next choose to fit Eq.(\ref{ALRphen}) to real beam asymmetry data \cite{CLAS:2015bqi}.  The results of the fit can be found in Figure \ref{ALRfit}.  Since the subleading twist term $\alpha_2$ is not expected to be small, we should include it in the fit.  However, leaving in the fit parameter $\beta_2$ leads to undesired results.  The low statistics of the data allows the possibility of generating a fit with singularities (vertical asymptotes) between data points when $\beta_2$ is included, which is relatively small compared to $\beta_1$ as it is.  Therefore, the best parametrization choice for this particular data is the expression
\begin{equation}\label{eq:ALRfinal}
    A_{LR}\approx\frac{\alpha_1\sin\phi+\alpha_2\sin(2\phi)}{1+\beta_1\cos\phi}\;.
\end{equation}

The results of the nonlinear fit in Figure \ref{ALRfit} give the parameter estimates: $\alpha_1=0.26\pm0.04$, $\alpha_2=0.07\pm0.17$ and $\beta_1=0.26\pm1.66$.  This suggests that this data is only useful to constrain the $\alpha_1$ parameter, but not the secondary $\alpha_2$ and $\beta_1$.  The question then arises: how precise must the data be in order to determine the secondary parameters in Eq. (\ref{eq:ALRfinal})?  To help answer this, we generate various pseudo-data for $A_{LR}(\phi)$ at the same kinematical point, varying the statistics and uncertainty.

To do so, we define 3 variables to modify the pseudo-data: $N$ for the number of evenly spaced angles $\phi$ assumed by the data points, $f$ for the maximum fractional fluctuation of the data points (which is randomized point-by-point) and $\delta$ for the fractional uncertainty of the data points themselves (with random fluctuations on the order of $\pm1/3$). Since we are interested in finding the harmonic structure of Eq. (\ref{eq:ALRfinal}) from such data, we assume that form in its generator.  For instance, the pseudo-data generator used for the kinematical point of the data from Figure \ref{ALRfit} is given by:
\begin{equation}\label{pseudoALR}
    A_{LR}^{\text{pseudo}}
    = \Bigg\{\frac{.25\sin\phi_i + 0.1\sin(2\phi_i)}{1+0.3\cos\phi_i} + \text{Rnd}[-0.25f,0.25f]\Bigg\} \pm \text{Rnd}\Bigg[0.25\bigg(\delta-\frac{\delta}{3}\bigg),0.25\bigg(\delta+\frac{\delta}{3}\bigg)\Bigg]\;,
\end{equation}
where $i=1,2,\cdots, N$ and $\text{Rnd}[x_1,x_2]$ generates a random number between $x_1$ and $x_2$.  

\begin{figure}[ht]
\centering
\includegraphics[width=0.75\textwidth]{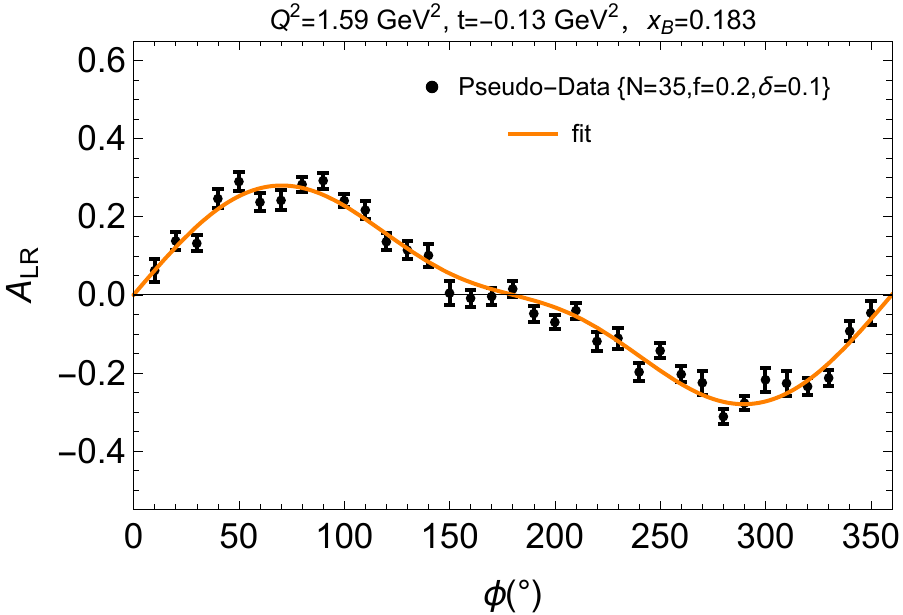}
\caption{\label{ALRpseudoFIT} The result of fitting Eq.(\ref{eq:ALRfinal}) to the pseudo-data generated by Eq. (\ref{pseudoALR}) with $N=35$, $f=0.2$ and $\delta=0.1$.}
\end{figure}

An example of the data points generated by Eq. (\ref{pseudoALR}) with a fitted Eq.(\ref{eq:ALRfinal}) is shown in Figure \ref{ALRpseudoFIT}.  There one can see a trace of the other harmonics around $\phi=\pi$.  By systematically varying $N$, $f$ and $\delta$, we find that the most important factor in determining a quality fit to $\alpha_2$ is $\delta$.  The difference in the fit between data with $N=17$ and $N=35$ for instance is small.  The effects of varying $f$ over $0.1<f<0.3$ was also not significant on the fit results.  Fixing $N=17$ and $f=0.1$ (reasonable expected values) while varying $\alpha_2$ from 5\% to 20\% revealed that 
\begin{equation}
    \frac{\Delta \alpha_2}{\alpha_2} \approx 2\frac{\Delta A_{LR}}{A_{LR}}\;,
\end{equation}
i.e. that the fractional uncertainty in the extracted $\alpha_2$ is approximately twice that of the fractional uncertainty of the data points of $A_{LR}$.  Therefore, in order to determine $\alpha_2$ to within 10\%, one should strive to have measured $A_{LR}$ to within 5\%, for example.  

Future beam spin asymmetry studies should attempt to include $\alpha_2$ as once one includes a twist-3 CFF analysis, this particular coefficient contains a twist-3 GPD associated with the OAM of quarks inside the proton \cite{Courtoy2014}.  Lastly, it appears safe to say that extracting $\beta_1$ with much precision, even from precision data, is unexpected.  Even for the pseudo-data shown in Figure \ref{ALRpseudoFIT}, $\beta_1$ cannot be determined better than to within 50\%.  We conclude that the observable $A_{LR}$ is not a reliable means to extract the real parts of the CFFs.

\subsection{CFF Extraction from Pseudo-data}

We will now study how to locally extract all 8 twist-2 CFFs using model-dependent pseudo-data for all of the total polarized and unpolarized cross sections covered in the text.  Since we do not yet possess real data for all of these potential observables, it will be instructive to learn what we can extract in the meantime with hypothetical data.  By using a particular GPD model, we will go in with a knowledge of our CFFs, allowing a benchmarking environment of the extraction method.

\begin{figure}[ht]
\centering
\begin{minipage}[b]{\textwidth}
\includegraphics[width=0.5\textwidth]{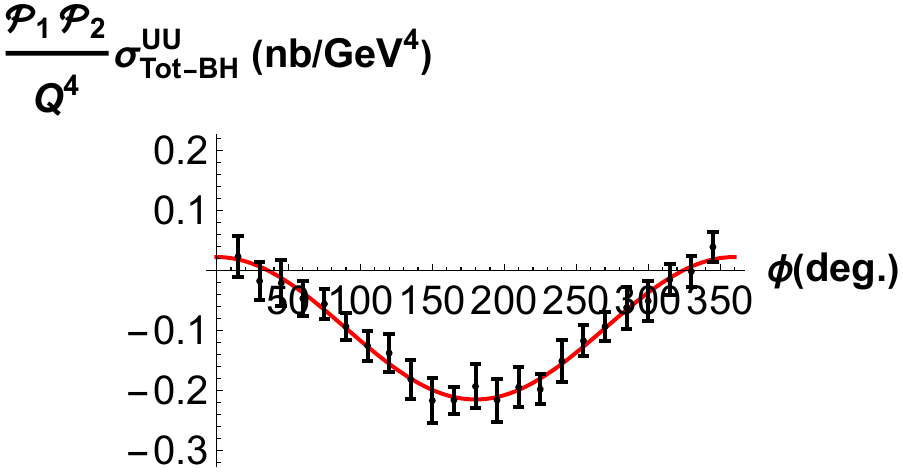}
\includegraphics[width=0.5\textwidth]{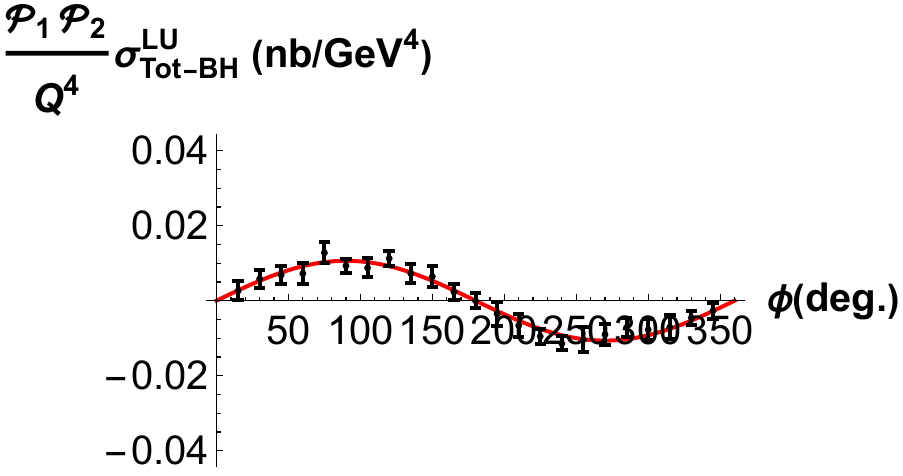}
\end{minipage}
\begin{minipage}[b]{\textwidth}
\includegraphics[width=0.5\textwidth]{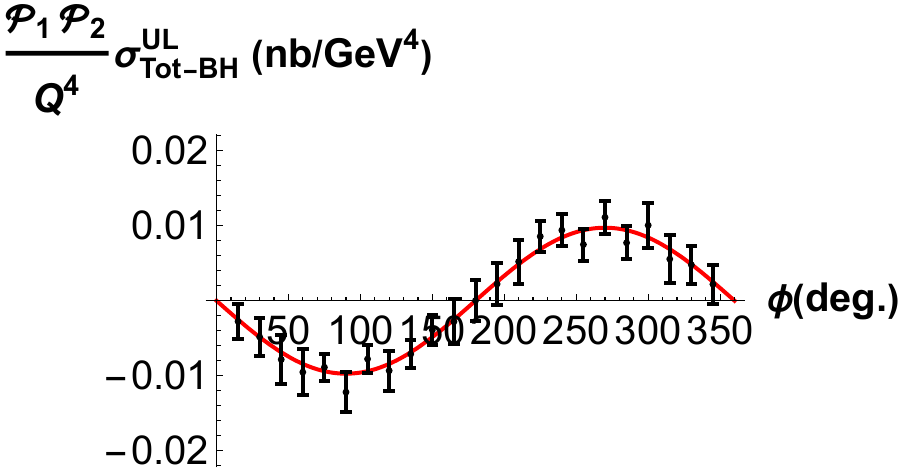}
\includegraphics[width=0.5\textwidth]{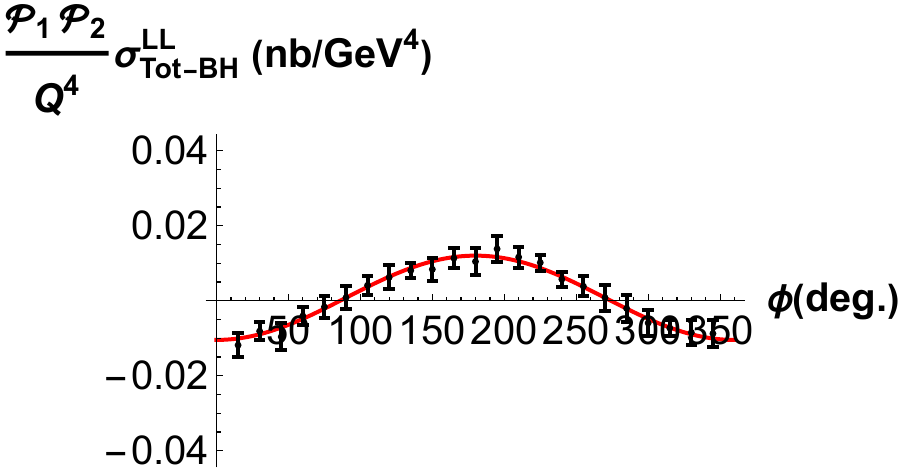}
\end{minipage}
\begin{minipage}[b]{\textwidth}
\includegraphics[width=0.5\textwidth]{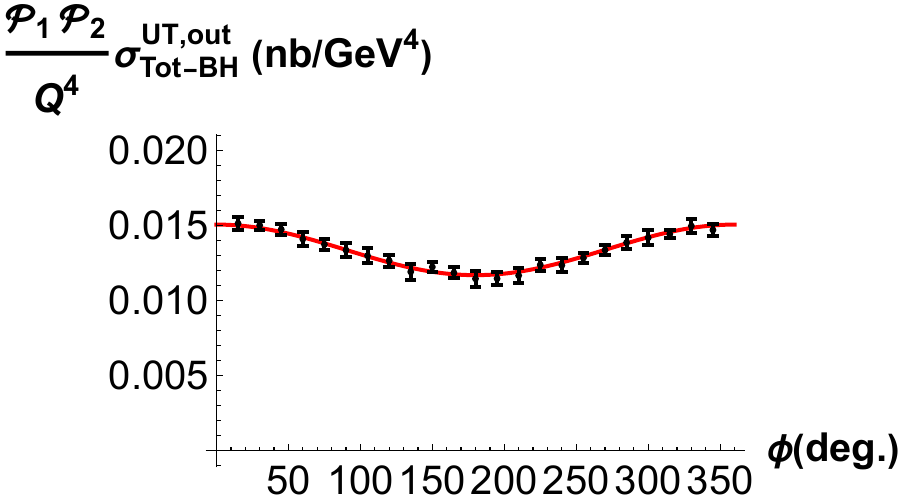}
\includegraphics[width=0.5\textwidth]{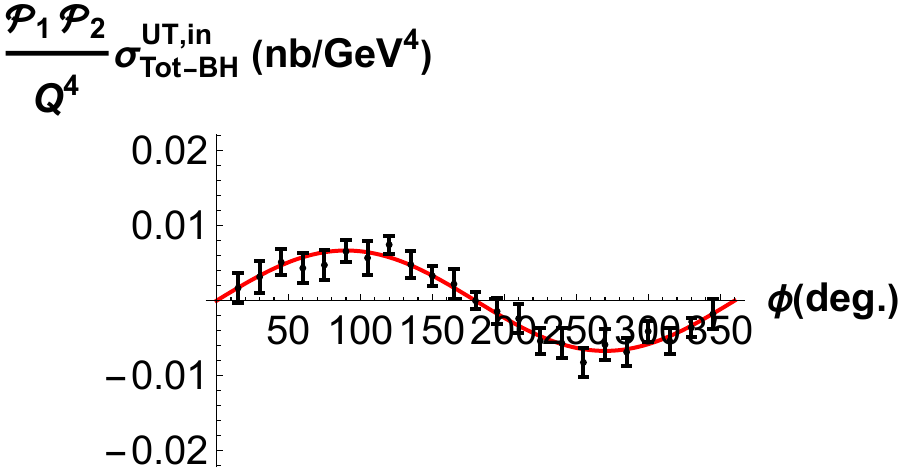}
\end{minipage}
\begin{minipage}[b]{\textwidth}
\includegraphics[width=0.5\textwidth]{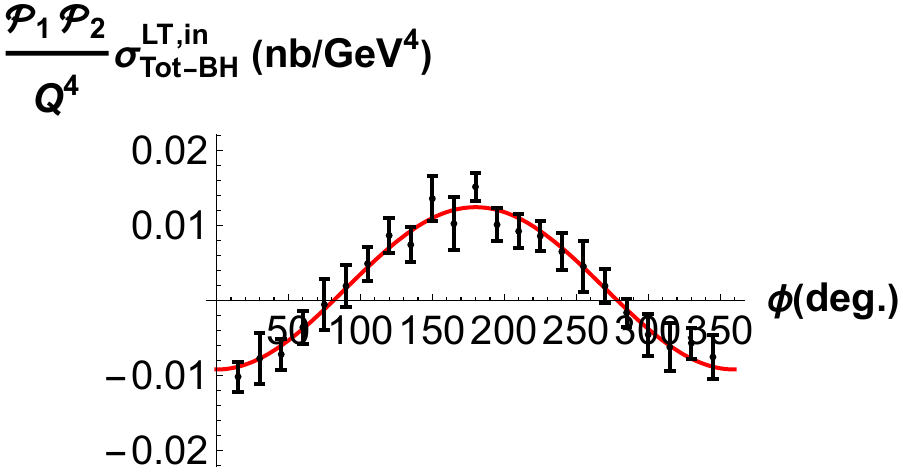}
\includegraphics[width=0.5\textwidth]{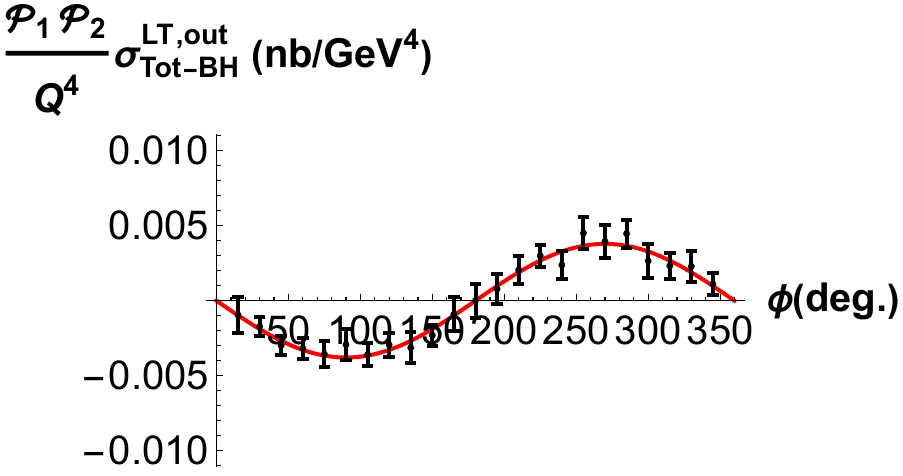}
\end{minipage}
\caption{\label{fig:pseudoplots} 8 generated DVCS pseudo-cross sections at $E_b=10.6~\text{GeV}$ using the data in Table \ref{tab:CFFmodel}.  All polarized cross sections have been taken with their helicity at $+1/2$, which is equivalent to taking $(\sigma^+-\sigma^-)/2$. The $UT$ and $LT$ cross sections have been integrated over  $\frac{1}{\pi}\int_0^{2\pi}\Delta\phi$ angle.  The data points and their uncertainties do not attempt to realistically represent real data at any particular facility, but merely use randomized statistical uncertainties $\sim 25\%$.}
\end{figure}

We begin by choosing one particular kinematical point: $Q^2=2.3~\text{GeV}^2,x_B=0.36, t=-0.28~\text{GeV}^2$, for which we use the determined value of the twist-2 CFFs using the UVa-based Reggeized di-quark model \cite{Gonzalez-Hernandez:2012xap,Kriesten:2021sqc} with origins from \cite{LiutiModel2011}, shown in Table \ref{tab:CFFmodel}.  Using these CFF values, we may predict a value for any DVCS cross section we like at any chosen value of $(E_b,\phi)$.  We shall fix the beam energy to $E_b=10.6~\text{GeV}$ and sample "measurement points" over discrete values of $\phi\in (0,2\pi)$.  This will be done in the same way as we did for $A_{LR}$ in Eq.(\ref{pseudoALR}), which involves randomized noise fluctuations of order $f$ percent and a percent uncertainty of order $\delta$.
\begin{table}
\centering
\scalebox{0.85}{
\begin{tabular}{|c|c|c|c|c|c|c|c|c|c|c|c|}
\hline
Model & $Q^2(\text{GeV}^2)$ & $x_B$ & $t(\text{GeV}^2)$ & Re$\mathcal{H}$ & Im$\mathcal{H}$ & Re$\mathcal{E}$ & Im$\mathcal{E}$ & Re$\widetilde{\mathcal{H}}$ & Im$\widetilde{\mathcal{H}}$ & Re$\widetilde{\mathcal{E}}$ & Im$\widetilde{\mathcal{E}}$ \\
\hline
UVa & $2.3$ & $0.36$ & $-0.28$ & $-0.7806$ & $1.951$ & $-0.3645$ & $0.709$ & $3.030$ & $1.044$ & $2.827$ & $5.331$ \\
\hline
\end{tabular}}
\caption{Proton CFF values used for generating pseudo-data using Reggeized diquark GPD model \cite{Gonzalez-Hernandez:2012xap,Kriesten:2021sqc}.} 
\label{tab:CFFmodel}
\end{table}
Making the choice of $f=0.3,~\delta=0.25$ all 8 cross sections $\frac{\mathcal{P}_1\mathcal{P}_2}{Q^4}\sigma^{P_bP_t}_{\text{Tot-BH}}(\phi)$ are plotted in Figure \ref{fig:pseudoplots} together with a fitted 1st-order harmonic function in $\phi$.  Although there are 8 cross measured cross sections here, 4 of them are of the form $a+b\cos\phi$ while the other 4 are fitted to $c\sin\phi$, totaling 12 fitted harmonic coefficients which will be used to constrain the twist-2 CFFs.  The uncertainty of the fitted coefficients is of order $\sim 10-20\%$.

The fitted values of all the 12 harmonic coefficients are then appropriately equated to Eqs.(\ref{sigUUapprox},\ref{sigLUapprox},\ref{sigULapprox},\ref{sigLLapprox},\ref{sigUTcosapprox},\ref{sigUTsinapprox},\ref{sigLTcosapprox},\ref{sigLTsinapprox}).  The resulting system of equations is indeed in the form of Eq.(\ref{8by8CFFsystem}).  Algebraically solving this system must not be done in the literal sense, but rather a numerical least-squares fit of the 8 unknown parameters needs to be performed to minimize the $\chi^2$ function. In this test, the minimized function is 
\begin{equation}\label{chisqgeneral}
    \chi^2=\sum_{i=1}^N\frac{(O_i-E_i)^2}{\sigma_i^2},
\end{equation}
where $N$ is the total number of experimental constraints, the observables $O_i$ are the fitted harmonic coefficients, $\sigma_i$ are their uncertainties, and the expected values $E_i$ are the predicted modified cross section harmonic coefficients, which depend on the unknown CFFs.
A numerical minimization of Eq.(\ref{chisqgeneral}) is performed using the optimal choice amongst various nonlinear methods including: Nelder Mead, differential evolution, stimulated annealing and random search; case-by-case.

\begin{figure}[ht]
\centering
\includegraphics[width=0.85\textwidth]{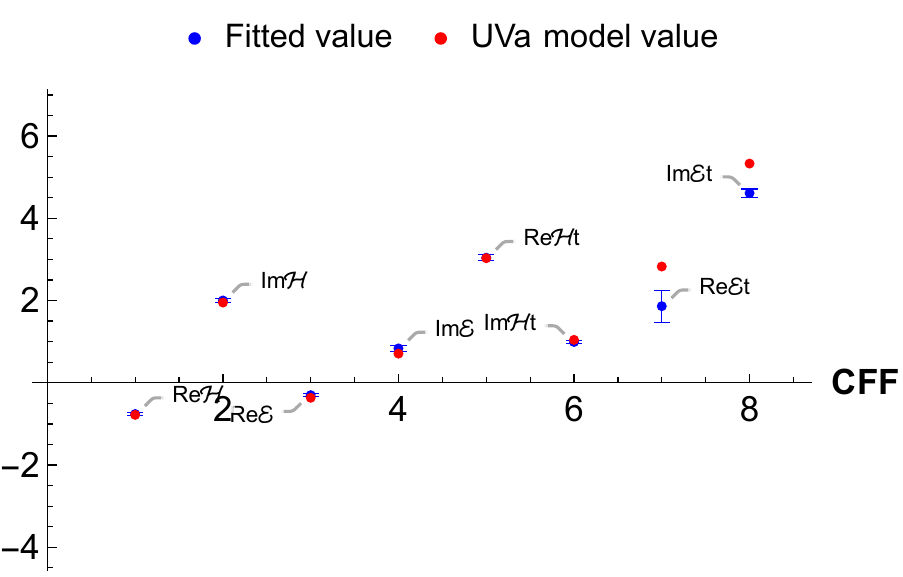}
\caption{\label{fig:extractedCFFsplot} Comparison of the 8 extracted twist-2 CFFs and their UVa model value used to generate the pseudo-data.}
\end{figure}

The results of this numerical least-squares minimization of $\chi^2$ can be found in Table \ref{tab:extractedCFFs} and Figure \ref{fig:extractedCFFsplot}.  The reduced $\chi^2_\nu=\chi^2/\nu$, where $\nu$ = the degrees of freedom = (\# of observables) $-$ (\# of fitting parameters) is found to be $0.21$.  The effect of introducing parameter constraints on the least-squares fit was investigated.  Similar as in \cite{Guidal2008}, we introduced a maximum threshold deviation from the model-value of the CFFs into the fit as a constraint.  For a threshold factor of $h$, the CFFs were constrained to remain within the domain
\begin{equation}
    \mathcal{F}_\text{model}-|h\mathcal{F}_\text{model}|\leq \mathcal{F} \leq \mathcal{F}_\text{model}+|h\mathcal{F}_\text{model}|\;,
\end{equation}
where the $\mathcal{F}_\text{model}$ are given in Table \ref{tab:CFFmodel}.  It was found that both the fitted value of the CFFs as well as $\chi^2_\nu$ were relatively insensitive to the value of $h\in [5,75)$.  We chose not to go below 5 so as to stay as model-independent as possible, while choosing thresholds above 75 resulted in undesired new parameter minima in the $\chi^2$.

A close inspection of all 12 equations for the harmonic coefficients involved reveals no apparent degeneracies, neither analytically nor numerically at the kinematical point chosen.  We therefore do not have any suspected grounds to remove any specific constraints from the fit on the concerns of a degeneracy.  One important question also is how important transversely-polarized target observables are towards an extraction of $\mathcal{E}$ specifically.  Removing the inclusion of any transversely-polarized target observables from the fit reduces the number of constraints to 6, resulting in $\nu=6-8<0$, which is not ideal.  We could, however, test whether a $UT$ or a $LT$ cross section would have a better effect on the extraction of $\mathcal{E}$.  The least-squares fit was repeated for either case and it was found that removing $UT$ measurements resulted in a better $\chi^2_\nu$ but large percent deviations in $\mathcal{E}$ with the UVa values, while removing $LT$ measurements gave a $\chi^2_\nu$ farther from 1, but with much more agreeable values of $\mathcal{E}$. 

\begin{table}
\centering
\scalebox{0.85}{
\begin{tabular}{|c|c|c|c|}
\hline
CFF & fit value & uncertainty & percent deviation \\
\hline
Re$\mathcal{H}$ & $-0.76$ & $0.04$ & $-3\%$ \\
\hline
Im$\mathcal{H}$ & $2.00$ & $0.05$ & $+3\%$ \\
\hline
Re$\mathcal{E}$ & $-0.30$ & $0.04$ & $+17\%$ \\
\hline
Im$\mathcal{E}$ & $0.84$ & $0.07$ & $+18\%$ \\
\hline
Re$\widetilde{\mathcal{H}}$ & $3.04$ & $0.07$ & $0\%$ \\
\hline
Im$\widetilde{\mathcal{H}}$ & $0.99$ & $0.03$ & $-5\%$ \\
\hline
Re$\widetilde{\mathcal{E}}$ & $1.86$ & $0.39$ & $-34\%$ \\
\hline
Im$\widetilde{\mathcal{E}}$ & $4.61$ & $0.10$ & $-14\%$ \\
\hline
\end{tabular}}
\caption{Fitted CFF values from $\chi^2$ minimization.  The percent deviation is from the expected UVa model values shown in Table \ref{tab:CFFmodel}. The uncertainties are computed via Eq.(\ref{eq:CFFunc}).} 
\label{tab:extractedCFFs}
\end{table}

We can still ask whether any specific subset of the 12 constraints gives an equivalent or improved CFF fit. With the insistence of maintaining at least $1$ degree of freedom, there are in principle $12!/(9!3!)=220$ ways to choose 9 constraints out of the 12 available.  Rather than exhaust all of these possible choices, we can systematically repeat fits with the removal of $UU,~LU,~UL~\text{and}~LL$ cross sections, and see if our $\chi^2_\nu$ and/or extracted CFF values improves.  Doing so reveals no particular improvement to the fit, with the exception of perhaps removing $LL$ cross sections, but with that choice only minor improvements are achieved.  All possible subsets of the 12 constraints leaves the extracted CFF values relatively unchanged with the exception of removing $UT$, which gives an unexpectedly small value of $\text{Re}\mathcal{E}$.  The maximum possible $\chi^2_\nu=0.21$ is achieved by keeping all 12 constraints in the fit.

The uncertainty for each of the fitted parameters is traditionally determined from the 2nd order derivative of the $\chi^2$ function with respect to that particular parameter via
\begin{equation}\label{eq:CFFunc}
    \Delta\mathcal{F}=\sqrt{2\Bigg( \frac{\partial^2\chi^2}{\partial\mathcal{F}^2}\Bigg)^{-1}}\;,
\end{equation}
which assumes that there are no correlations between each parameter. We find that this may in fact underestimate the uncertainty of the extracted CFFs. This can be seen by looking at how much the fitted values differ from the model values in Figure \ref{fig:extractedCFFsplot}, and these are quantified in values ranging from $0-34\%$ in Table \ref{tab:extractedCFFs}.  Surely, further studies of how error propagation should ensue between the cross section data and the extracted CFFs is warranted here, and is left for a future work.

So far we have been applying a $\chi^2$ fit to the harmonic coefficients.  Alternatively, one can in fact use cross sections directly in Eq.(\ref{chisqgeneral}).  This would negate the need to perform harmonic fits to the data while substantially increasing the number of degrees of freedom in the fit.  We have performed this exercise on all of the data points shown in Figure \ref{fig:pseudoplots}.  The result is nearly identical central values of the extracted CFFs and a nearly identical $\chi^2_\nu$ but with somewhat smaller CFF uncertainties after the application of Eq.(\ref{eq:CFFunc}).  We advocate the step of harmonic fitting, as it helps one organize the inclusion/exclusion of higher-twist contributions.  Fitting the CFFs to harmonic coefficients should also in principle simplify the location of the global minimum in the $\chi^2$, as having many more terms in Eq.(\ref{chisqgeneral}) will have a much more complicated geometry in its hyper-dimensional 8-space. 

\section{Conclusion}
\label{sec:5}

In this study, we looked at the exercise of locally extracting the twist-2 CFFs from DVCS observables.  In order to do this, we computed all possible polarized/unpolarized DVCS total cross sections, as well as the beam spin asymmetry, in terms of the twist-2 CFFs explicitly.  Although we possess all-orders of the kinematics, for practical purposes we only provide all of the lab frame kinematics accurate up to twist-4 in this paper.  Furthermore, we choose to decompose all of our observables into harmonic form (i.e. in $\cos(n\phi)$ and $\sin(n\phi)$).

The general exercise of locally extracting the eight assumed free variables\newline $\{\text{Re},\text{Im}\} \{\mathcal{H},\mathcal{E},\widetilde{\mathcal{H}},\widetilde{\mathcal{E}}\}(x_B,t,Q)$ from available DVCS observables is indeed possible, given enough data.  There is little doubt that the global fitting of CFFs as well as machine learning techniques will one day allow the extraction of the more elusive GPDs.  However, we stress the principle importance of having a sufficient number of constraints (at least 8 distinct measured observables) to determine the twist-2 CFFs, and that this model-independent approach will serve as a crucial guide for future techniques.  

To aid in the exploitation of the azimuthal dependence of the DVCS cross sections, we prescribe the definition of re-weighted cross sections, which involves the multiplication of the BH propagators $\mathcal{P}_1(\phi)\mathcal{P}_2(\phi)$.  Trigonometric identities then allows us to reduce these reduced cross sections to very simple $\phi$ expressions practical for CFF extraction.

We demonstrated that the DVCS cross sections may be greatly simplified in their harmonic structure by keeping only terms which includes leading twist kinematics, which resides in the lower frequency harmonic terms.  Those remaining coefficients are then weighted over expected JLab 12 GeV lab frame kinematics, identifying the dominant harmonic coefficients, simplifying the DVCS observables yet further.  From these results, we propose a roadmap for twist-2 CFF extraction from total DVCS cross sections alone, as opposed to prior attempts to extract solely from asymmetries.  We have also excluded the possibility of positron beams.  We demonstrate that one can in fact successfully extract the harmonic coefficients from real data, using both the unpolarized DVCS cross section and the beam spin asymmetry as examples. 

Using a specific GPD model \cite{LiutiModel2011}, we have generated pseudo-data for all possible DVCS cross sections with conservative uncertainties and statistics.  From this pseudo-data, we have placed 12 numerical constraints on the twist-2 CFFs through fitted harmonic coefficients.  A $\chi^2$ fit was then performed on those equations, and all eight CFFs were determined with error estimates. Various systematic studies were done on this fitting procedure.  The general conclusion is that this technique is successful, and using as many constraints as possible generally gives a better fit.  Understanding the final error estimates of the extracted CFFs is non-trivial, and warrants further investigation.

\section*{Acknowledgments}
We thank B. Kriesten, F-X. Girod, M. Burkardt and N. Sato for discussions related to the subject of this paper. This research is supported by the U.S. Department of Energy, Office of Science, Office of Nuclear Physics, under contract number DE-SC0020682, and the Center for Nuclear Femtography, Southeastern Universities Research Association, Washington D.C.

\newpage
\appendix

\section{Pure DVCS Polarization Cross Sections}
\label{App:A}

The pure DVCS cross section will be fully considered up to twist-2 dynamics in this section. To start, we may decompose the general pure DVCS cross section by polarization cases as follows
\begin{equation}
    \sigma_{\text{DVCS}}=\sigma_{\text{DVCS}}^{UU} +  \sigma_{\text{DVCS}}^{LL}(h,\Lambda_l) + \sigma_{\text{DVCS}}^{UT}(\Lambda_T) + \sigma_{\text{DVCS}}^{LT}(h,\Lambda_T)\;,
\end{equation}
where each of the 4 polarization cross sections is given below, and are linearly dependent on their helicity parameter(s). Note the $LU$ DVCS cross section here is identically zero simply due to its vanishing Dirac structure, while the $UL$ case only arises at twist-3.  The equations which will ensue in this appendix are taken directly from \cite{Guo:2021gru} with the added intention to both explicitly express the DVCS cross sections in terms of the real and imaginary parts of the twist-2 CFFs and provide the lab-frame kinematical coefficients explicitly through harmonic coefficients.

\subsection{Unpolarized Beam Unpolarized Target}
\label{sec:A1}

The unpolarized DVCS contribution to the cross section is given by

\begin{equation}\label{sigdvcsUU}
   \sigma^{UU}_{\text{DVCS}}=\frac{\Gamma}{Q^4}F^{UU}\;, 
\end{equation}

\begin{eqnarray}\label{FUU}
    F^{UU}=4h^U\Bigg{\{} (1-\xi^2)\big{[}(\text{Re}\mathcal{H})^2+(\text{Im}\mathcal{H})^2 +(\text{Re}\mathcal{\tilde{H}})^2+(\text{Im}\mathcal{\tilde{H}})^2 \big{]}\nonumber\\
    -\frac{t}{4M^2}\big{[} (\text{Re}\mathcal{E})^2+(\text{Im}\mathcal{E})^2 +\xi^2(\text{Re}\mathcal{\tilde{E}})^2+(\text{Im}\mathcal{\tilde{E}})^2 \big{]}\nonumber\\
    -\xi^2\Bigg{[}(\text{Re}\mathcal{E})^2+ (\text{Im}\mathcal{E})^2  +2\text{Re}\mathcal{E}\text{Re}\mathcal{H} + \text{Im}\mathcal{E}\text{Im}\mathcal{H} \nonumber\\
    +2\text{Re}\mathcal{\tilde{E}}\text{Re}\mathcal{\tilde{H}} + \text{Im}\mathcal{\tilde{E}}\text{Im}\mathcal{\tilde{H}} \Bigg{]}
    \Bigg{\}}\;,
\end{eqnarray}
involving the scalar quantity $h^U$, which is expressible in terms of an azimuthal-even harmonic series
\begin{eqnarray}
h^U&=&\sum_{n=0}^{3}h_n^U\cos(n\phi)\; .\label{hU}
\end{eqnarray}  
Eq.(\ref{FUU}) also involves the skewness parameter, which is given by
\begin{equation}
    \xi=\frac{x_B}{2-x_B}-\frac{2x_B(x_B-1)t}{(x_B-2)^2Q^2}\;,
\end{equation}
in our light cone choice.
We perform a twist expansion ($1/Q$-expansion) to the harmonic coefficients defined above up to $\text{N}^3$LO.  We caution the reader that these expansions in some cases may not be sufficiently accurate for a high precision study of DVCS data, in which case the unapproximated expressions should be used \cite{Guo:2021git}.  The twist-expanded coefficients presented here are merely used for the pragmatic purpose of determining the relative size of the various harmonic coefficients which enters the DVCS observables considered in the text. In the lab frame, the coefficients of Eq. (\ref{hU}) are given up to kinematic twist-4 accuracy by
 \begin{eqnarray}
 h_0^U&\approx& \frac{-4M^2x_B^2(y^2-5y+5)+Q^2(2-2y+y^2)+2t(x_B-1)(y^2-6y+6)}{2y^2}\; , \\
 h_1^U&\approx& \frac{2Q(y-2)\sqrt{[t(x_B-1)-M^2x_B^2](1-y)}}{y^2}\; ,\\
 h_2^U&\approx&-\frac{2(y-1)[t(x_B-1)-M^2x_B^2]}{y^2}\; ,\\
 h_3^U&\approx&0\; ,
 \end{eqnarray}
 in which we have taken the light cone direction associated with $\beta\rightarrow\infty$ in \cite{Guo:2021gru}.
 






\subsection{Polarized beam Longitudinally polarized Target}

The $LL$ DVCS cross section is given by
\begin{equation}\label{sigdvcsLL}
    \sigma^{LL}_{\text{DVCS}}=\frac{\Gamma}{Q^4}(2h)(2\Lambda_l)F^{LL}\;,
\end{equation}

\begin{eqnarray}\label{FLL}
    F^{LL}&=&-8h^{-,L}\Bigg{\{}
    (1-\xi^2)(\text{Re}\mathcal{\widetilde{H}}\text{Re}\mathcal{H}+\text{Im}\mathcal{\widetilde{H}}\text{Im}\mathcal{H}) \nonumber\\
    &~&-\xi^2(\text{Re}\mathcal{\widetilde{H}}\text{Re}\mathcal{E}+\text{Im}\mathcal{\widetilde{H}}\text{Im}\mathcal{E}+\text{Re}\mathcal{\widetilde{E}}\text{Re}\mathcal{H}+\text{Im}\mathcal{\widetilde{E}}\text{Im}\mathcal{H}) \nonumber\\
    &~&-\bigg{(}\frac{\xi^2}{1+\xi}+\frac{t}{4M^2} \bigg{)}\xi (\text{Re}\mathcal{\widetilde{E}}\text{Re}\mathcal{E}+\text{Im}\mathcal{\widetilde{E}}\text{Im}\mathcal{E})
    \Bigg{\}}\;,
\end{eqnarray}
in which the sole scalar coefficient is given by
\begin{equation}\label{hmL}
   h^{-,L}=\sum_{n=0}^{3}h_n^{-,L}\cos(n\phi) \;.
\end{equation}
The polarized scalar harmonic coefficients in Eq. (\ref{hmL}) are given to twist-4 accuracy in the lab frame by
 \begin{eqnarray}
 h_0^{-,L}&\approx&\frac{(y-2)[-4M^2x_B^2+Q^2+2tx_B-2t]}{2y}\; ,\\
 h_1^{-,L}&\approx&-\frac{2Q\sqrt{[t(x_B-1)-M^2x_B^2](1-y)}}{y}\; ,\\
 h_2^{-,L}&\approx&0\approx h_3^{-,L}\; .
 \end{eqnarray}
 
\subsection{Unpolarized beam Transversely polarized Target}

Suppose we have an unpolarized beam scattering off of a transversely polarized target with polarization vector
\begin{equation}\label{Svector}
    S^\mu = 2\Lambda_T(0,\cos\phi_S,\sin\phi_S,0)\;,
\end{equation}
where $\phi_S$ is the azimuthal angle of the target polarization from the lab frame $x$-axis about the positive $z$-axis.  Then the pure DVCS cross section will be given by
\begin{equation}\label{sigdvcsUT}
    \sigma_{\text{DVCS}}^{UT}= \sin(\phi_S-\phi)\sigma_{\text{DVCS}}^{UT,\rm{out}}\;.
\end{equation}
Here the other potential term in which the polarization vector of the proton is parallel to the hadronic plane is zero. 





\paragraph{Perpendicular to hadronic plane}

The DVCS cross section in which the target nucleon's polarization vector is perpendicular to the hadronic $(p',q')$ plane is given by
\begin{equation}\label{sigdvcsUtout}
   \sigma^{UT,\rm{out}}_{\text{DVCS}}=\frac{\Gamma}{Q^4}(2\Lambda_T)F^{UT,\rm{out}}\;, 
\end{equation}
\begin{eqnarray}\label{FUTout}
F^{UT,\text{out}}&=&4Nh^U\Bigg[ \text{Re}\mathcal{H}\text{Im}\mathcal{E} - \text{Re}\mathcal{E}\text{Im}\mathcal{H} - \xi(\text{Re}\mathcal{\widetilde{H}}\text{Im}\mathcal{\widetilde{E}}-\text{Re}\mathcal{\widetilde{E}}\text{Im}\mathcal{\widetilde{H}}) 
\Bigg]\;,
\end{eqnarray}
where $h^U$ may again may assume the form of Eq.(\ref{hU}) and the normalization factor introduced in \cite{Guo:2021gru} is given by
\begin{equation}
    N=\frac{\sqrt{-4M^2\xi^2-t(1-\xi^2)}}{M}\; .
\end{equation}

\subsection{Polarized beam Transversely polarized Target}

Now with a polarized beam we define the total pure DVCS $LT$ cross section similarly as we did the $UT$,
\begin{equation}\label{sigdvcsLT}
     \sigma_{\text{DVCS}}^{LT}=\cos(\phi_S-\phi)\sigma_{\text{DVCS}}^{LT,\rm{in}}\;,
\end{equation}
which now does not have a component perpendicular to the hadronic plane.

\paragraph{Parallel to hadronic plane}

The DVCS cross section in which the target nucleon's polarization vector is parallel to the hadronic $(p',q')$ plane is given by
\begin{equation}\label{sigdvcsLTin}
   \sigma^{LT,\rm{in}}_{\text{DVCS}}=\frac{\Gamma}{Q^4}(2h)(2\Lambda_T)F^{LT,\rm{in}}\;, 
\end{equation}
\begin{eqnarray}\label{FLTin}
F^{LT,\rm{in}}&=&-4Nh^{-,L}\Bigg[\text{Re}\mathcal{\widetilde{H}}\text{Re}\mathcal{E} + \text{Im}\mathcal{E}\text{Im}\mathcal{\widetilde{H}} - \xi(\text{Re}\mathcal{\widetilde{E}}\text{Re}\mathcal{H}+\text{Im}\mathcal{\widetilde{E}}\text{Im}\mathcal{H}) \nonumber\\
&~&\;\;\;\;\;\;\;\;\;\;-\frac{\xi^2}{1+\xi}(\text{Re}\mathcal{\widetilde{E}}\text{Re}\mathcal{E}+\text{Im}\mathcal{\widetilde{E}}\text{Im}\mathcal{E})
\Bigg]\;,
\end{eqnarray}
and again, $h^{-,L}$ may be decomposed into harmonic series of Eq. (\ref{hmL}).  It is important to note that for a proton transversely polarized perpendicular to the hadronic plane, there is no pure DVCS contribution to the total cross section. This concludes our pure DVCS cross sections.

\section{Interference Polarization Cross Sections}
\label{App:B}

The interference cross section is generally of a comparable magnitude to the pure DVCS one, but has a more rich azimuthal structure. 
Just as we did for the DVCS case, we choose to decompose this interference cross section by beam and target polarizations,
\begin{equation}
    \sigma_{\mathcal{I}}=\sigma_{\mathcal{I}}^{UU} + \sigma_{\mathcal{I}}^{UL}(\Lambda_l) + \sigma_{\mathcal{I}}^{LU}(h) + \sigma_{\mathcal{I}}^{LL}(h,\Lambda_l) + \sigma_{\mathcal{I}}^{UT}(\Lambda_T) + \sigma_{\mathcal{I}}^{LT}(h,\Lambda_T)\;,
\end{equation}
which now includes all 6 possible combinations of polarized and unpolarized beam and target.  Each of these cross sections is given explicitly in this section, once again in terms of the real and imaginary parts of the twist-2 CFFs and providing the lab-frame kinematical coefficients explicitly as determined from \cite{Guo:2021gru}. 

\subsection{Unpolarized Beam Unpolarized Target}
\label{sec:B1}

The $UU$ interference cross section for an electron/positron beam of charge $(e_l=\pm1)$ is given by
\begin{equation}\label{sigIUU}
   \sigma^{UU}_\mathcal{I} = -\frac{e_l\Gamma}{Q^2 t}F^{UU}_\mathcal{I}\;,
\end{equation}
\begin{eqnarray}\label{FIUU}
   F^{UU}_\mathcal{I}=A^{I,\rm{U}} \left(F_1 \text{Re}\mathcal {H}-\frac{t}{4M^2} F_2 \text{Re}\mathcal {E}\right)+B^{I,\rm{U}} (F_1+F_2)(\text{Re}\mathcal {H}+\text{Re}\mathcal {E}) \nonumber\\
+ C^{I,\rm{U}}(F_1+F_2){\text{Re}\mathcal {\widetilde{H}}}\; .
\end{eqnarray}
The benefit of using this form is twofold: first, the CFF dependence is explicit and second, the coefficients $A^{I,U},B^{I,U}$ and $C^{I,U}$ can be calculated to a high kinematical precision. This form of the interference cross section was introduced in \cite{Kriesten:2019jep}, wherein physical connections are made to Rosenbluth separation and inclusive processes. 

We will also need to fully understand the azimuthal dependence of the cross section, and to help us do so, we will expand these 3 coefficients into a harmonic series as follows
\begin{eqnarray}
    A^{I,U}=\frac{Q^4}{\mathcal{P}_1(\phi)\mathcal{P}_2(\phi)}\sum_{n=0}^{3}a_n^{I,U}\cos(n\phi)\;,\label{AIU}\\
     B^{I,U}=\frac{Q^4}{\mathcal{P}_1(\phi)\mathcal{P}_2(\phi)}\sum_{n=0}^{3}b_n^{I,U}\cos(n\phi)\;,\label{BIU}\\
      C^{I,U}=\frac{Q^4}{\mathcal{P}_1(\phi)\mathcal{P}_2(\phi)}\sum_{n=0}^{3}c_n^{I,U}\cos(n\phi)\;,\label{CIU}
\end{eqnarray}
 The factor $\mathcal{P}_1\mathcal{P}_2$ is given explicitly as a function of $(x_B,Q,t,\phi,y)$ in Eqs.(\ref{P1P2harm}-\ref{BH2}) The harmonic coefficients in Eqs.(\ref{AIU}-\ref{CIU}) are given to twist-4 accuracy in the lab frame below.
 
 \paragraph{$A^{I,U}$ Coefficients:}
\begin{eqnarray}
a_3^{I,U}&\approx&-\frac{16\kappa_t^\frac{3}{2}}{Qx_By^3}\; ,\\
a_2^{I,U}&\approx&-\frac{16(y-2)\kappa_t}{x_By^3}\; ,\\
a_1^{I,U}&\approx&-\frac{8\kappa_t^\frac{1}{2}}{Qx_By^3}\bigg[ -2M^2x_B^2(4y^2-19y+19) +Q^2(y^2-2y+2) +2tx_B(3y^2-11y+11)\nonumber\\
&~&\; -4t(y^2-3y+3) \bigg]\; ,\\
a_0^{I,U}&\approx&\frac{8(y-2)}{x_By^3}[M^2x_B^2(y-2)^2+t\big( -x_B(y^2-3y+3)+y^2-2y+2 \big)]\; ,
\end{eqnarray}
which involve the recurring kinematical factor responsible for the minimum allowed magnitude of $t$,
\begin{equation}
  \kappa_t\equiv (1-y)[t(x_B-1)-M^2x_B^2]\; .  
\end{equation}

\paragraph{$B^{I,U}$ Coefficients:}
\begin{eqnarray}
b_3^{I,U}&\approx&0\approx b_2^{I,U}\; ,\\
b_1^{I,U}&\approx&-\frac{16tx_B(y^2-6y+6)\kappa_t^\frac{1}{2}}{Q(x_B-2)y^3}\; ,\\
b_0^{I,U}&\approx&\frac{8tx_B(y^2-3y+2)}{(x_B-2)y^3}\; ,
\end{eqnarray}
\paragraph{$C^{I,U}$ Coefficients:}
\begin{eqnarray}
c_3^{I,U}&\approx&-\frac{x_B}{x_B-2}a_3^{I,U}\; ,\\
c_2^{I,U}&\approx&-\frac{x_B}{x_B-2}a_2^{I,U}\; ,\\
c_1^{I,U}&\approx&-\frac{x_B}{x_B-2}a_1^{I,U}+\frac{16t\kappa_t^\frac{1}{2}}{Q(x_B-2)^2y^3}\bigg\{ x_B^2(y^2-6y+6)+x_B(-3y^2+20y-20)+3(y^2-6y+6)\bigg\}\; ,\nonumber\\
&~&\\
c_0^{I,U}&\approx&-\frac{8(y-2)^3\kappa_t}{(x_B-2)y^3(1-y)}\; .
\end{eqnarray}

\subsection{Unpolaried Beam Longitudinally Polarized Target}

The total interference cross section for an unpolarized beam on a longitudinally polarized target ($\Lambda_l=\pm1/2$) is given by
\begin{equation}\label{sigIUL}
   \sigma^{UL}_\mathcal{I} = -\frac{e_l\Gamma}{Q^2 t}(2\Lambda_l)F^{UL}_\mathcal{I}\;,
\end{equation}
\begin{eqnarray}\label{FIUL}
   F^{UL}_\mathcal{I}=\widetilde{A}^{I,\rm{U}}\Bigg[ F_1 \left(\text{Im}\mathcal {\widetilde{H}}-\frac{\xi^2}{1+\xi} \text{Im}\mathcal {\widetilde{E}}\right)-F_2\frac{t}{4M^2}\text{Im}\mathcal{\widetilde{E}}\Bigg]\nonumber\\
   +\widetilde{B}^{I,\rm{U}} (F_1+F_2)\left(\text{Im}\mathcal {\widetilde{H}}+\frac{\xi}{1+\xi}\text{Im}\mathcal {\widetilde{E}}\right) \nonumber\\
+ \widetilde{C}^{I,\rm{U}}(F_1+F_2) \left( {\text{Im}\mathcal {H}}+\frac{\xi}{1+\xi}\text{Im}\mathcal{E}\right)\; .
\end{eqnarray}
Once again, we may expand the interference coefficients into a harmonic series, this time in terms of trigonometric sine functions
\begin{eqnarray}
    \tilde{A}^{I,U}=\frac{Q^4}{\mathcal{P}_1(\phi)\mathcal{P}_2(\phi)}\sum_{n=1}^{3}\tilde{a}_n^{I,U}\sin(n\phi)\;,\label{AtIU}\\
     \tilde{B}^{I,U}=\frac{Q^4}{\mathcal{P}_1(\phi)\mathcal{P}_2(\phi)}\sum_{n=1}^{3}\tilde{b}_n^{I,U}\sin(n\phi)\;,\label{BtIU}\\
      \tilde{C}^{I,U}=\frac{Q^4}{\mathcal{P}_1(\phi)\mathcal{P}_2(\phi)}\sum_{n=1}^{3}\tilde{c}_n^{I,U}\sin(n\phi)\;,\label{CtIU}
\end{eqnarray}
whose coefficients are distinguished by those of Eqs. (\ref{AIU})-(\ref{CIU}) by the presence of a tilde, and are given below.

\paragraph{$\widetilde{A}^{I,U}$ Coefficients:}

\begin{eqnarray}
\widetilde{a}_3^{I,U}&\approx&\frac{16\kappa_t^\frac{3}{2}}{Qx_By^3}\; , \\
\widetilde{a}_2^{I,U}&\approx&-\frac{16(y-2)\kappa_t}{x_By^3}\; ,\\
\widetilde{a}_1^{I,U}&\approx& \frac{4\kappa_t^{-\frac{1}{2}}}{Qx_By^3} \bigg\{ 2Q^2(y^3-3y^2+4y-2)(M^2x_B^2-tx_B+t)
+tM^2x_B^2\big[ -x_B(y^4-18y^3+82y^2\nonumber\\
&~&\;-128y+64)+y^4-16y^3+72y^2-112y+56 \big]+M^4x_B^4\big[y^4 -14y^3+58y^2-88y+44 \big]\nonumber\\
&~&\; -t^2(x_B-1)(y-1)\big[ x_B(5y^2-22y+22)-4(y-2)^2 \big]\bigg\}\; ,\\
\widetilde{a}_0^{I,U}&\approx&0\; ,
\end{eqnarray}

\paragraph{$\widetilde{B}^{I,U}$ Coefficients:}
 \begin{eqnarray}
 \widetilde{b}_3^{I,U}&\approx&0\approx \widetilde{b}_2^{I,U} \approx \widetilde{b}_1^{I,U}\; ,
 \end{eqnarray}
 
\paragraph{$\widetilde{C}^{I,U}$ Coefficients:}

\begin{eqnarray}
\widetilde{c}_3^{I,U}&\approx&\frac{x_B}{x_B-2}\widetilde{a}_3^{I,U}\; ,\\
\widetilde{c}_2^{I,U}&\approx&\frac{x_B}{x_B-2}\widetilde{a}_2^{I,U}\; ,\\
\widetilde{c}_1^{I,U}&\approx&\frac{x_B}{x_B-2}\widetilde{a}_1^{I,U} -\frac{16t\kappa_t^\frac{1}{2}[(x_B-1)y^2-2y+2]}{Q(x_B-2)^2y^3}\; , \\
\widetilde{c}_0^{I,U}&\approx&0\; .
\end{eqnarray}

\subsection{Polarized Beam Unpolarized Target}

The total interference cross section for a polarized beam ($h=\pm1/2$) on an unpolarized target  is given by
\begin{equation}\label{sigILU}
   \sigma^{LU}_\mathcal{I} = -\frac{e_l\Gamma}{Q^2 t}(2h)F^{LU}_\mathcal{I},
\end{equation}
\begin{eqnarray}\label{FILU}
  F_{\mathcal{I}}^{LU}=A^{I,\rm{L}} \left(F_1 \text{Im}\mathcal {H}-\frac{t}{4M^2} F_2 \text{Im}\mathcal {E}\right)&+&B^{I,\rm{L}} (F_1+F_2)(\text{Im}\mathcal {H}+\text{Im}\mathcal {E}) \\
    &\;&+
     C^{I,\rm{L}}(F_1+F_2)\text{Im}\widetilde{\mathcal {H}}\; .
\end{eqnarray}
The coefficients of which are given by
\begin{eqnarray}
    A^{I,L}=\frac{Q^4}{\mathcal{P}_1(\phi)\mathcal{P}_2(\phi)}\sum_{n=1}^{3}a_n^{I,L}\sin(n\phi)\;,\label{AIL}\\
     B^{I,L}=\frac{Q^4}{\mathcal{P}_1(\phi)\mathcal{P}_2(\phi)}\sum_{n=1}^{3}b_n^{I,L}\sin(n\phi)\;,\label{BIL}\\
      C^{I,L}=\frac{Q^4}{\mathcal{P}_1(\phi)\mathcal{P}_2(\phi)}\sum_{n=1}^{3}c_n^{I,L}\sin(n\phi)\;.\label{CIL}
\end{eqnarray}
These polarized scalar harmonic coefficients are given below.

\paragraph{$A^{I,L}$ Coefficients:}

\begin{eqnarray}
a_3^{I,L}&\approx&0\; ,\\
a_2^{I,L}&\approx&\frac{16\kappa_t}{x_By^2}\; ,\\
a_1^{I,L}&\approx&\frac{4(y-2)\kappa_t^{-\frac{1}{2}}}{Qx_By^2} \bigg\{ M^2tx_B^2[-x_B(y^2-16y+16)+y^2-14y+14] +M^4x_B^4(y^2-12y+12)\nonumber\\
&~&\; -t^2(5x_B^2-9x_B+4)(y-1) -2Q^2\kappa_t \bigg\}\; ,\\
\end{eqnarray}

\paragraph{$B^{I,L}$ Coefficients:}

\begin{eqnarray}\label{bILharmonics}
b_3^{I,L}&\approx&0\approx b_2^{I,L}\approx b_1^{I,L}\; ,
\end{eqnarray}

\paragraph{$C^{I,L}$ Coefficients:}

\begin{eqnarray}
c_3^{I,L}&\approx&0\; ,\\
c_2^{I,L}&\approx&-\frac{x_B}{x_B-2}a_2^{I,L}\; ,\\
c_1^{I,L}&\approx&-\frac{x_B}{x_B-2}a_1^{I,L} -\frac{16t(x_B-1)(y-2)\kappa_t^\frac{1}{2}}{Q(x_B-2)^2y^2}\; .
\end{eqnarray}

\subsection{Polarized beam Longitudinally polarized Target}

For the interference case of a polarized beam and longitudinally polarized target we have the cross section
\begin{equation}\label{sigILL}
   \sigma^{LL}_\mathcal{I} = -\frac{e_l\Gamma}{Q^2 t}(2h)(2\Lambda_l)F^{LL}_\mathcal{I}\;,
\end{equation}
\begin{eqnarray}\label{FILL}
 F_{\mathcal{I}}^{LL}&=& \tilde A^{I,\rm{L}}\left[F_1\left(\text{Re}\widetilde{ \mathcal {H}}-\frac{\xi^2}{1+\xi}\text{Re}\widetilde{ \mathcal {E}} \right)-F_2\frac{t}{4M^2}\xi\text{Re}\widetilde{ \mathcal {E}}\right]\\
       &+&\tilde B^{I,\rm{L}} (F_1+F_2)\left(\text{Re}\widetilde{ \mathcal {H}}+\frac{\xi}{1+\xi} \text{Re}\widetilde{\mathcal {E}} \right) \\ &-&\tilde C^{I,\rm{L}}(F_1+F_2)\left( \text{Re}\mathcal {H}+\frac{\xi}{1+\xi} \text{Re}\mathcal {E} \right)\; .
\end{eqnarray}
These double-spin coefficients above relegate back to an azimuthal-even harmonic series of cosines 
\begin{eqnarray}
    \tilde{A}^{I,L}=\frac{Q^4}{\mathcal{P}_1(\phi)\mathcal{P}_2(\phi)}\sum_{n=0}^{3}\tilde{a}_n^{I,L}\cos(n\phi)\;,\label{AtIL}\\
     \tilde{B}^{I,L}=\frac{Q^4}{\mathcal{P}_1(\phi)\mathcal{P}_2(\phi)}\sum_{n=0}^{3}\tilde{b}_n^{I,L}\cos(n\phi)\;,\label{BtIL}\\
      \tilde{C}^{I,L}=\frac{Q^4}{\mathcal{P}_1(\phi)\mathcal{P}_2(\phi)}\sum_{n=0}^{3}\tilde{c}_n^{I,L}\cos(n\phi)\;,\label{CtIL}
\end{eqnarray}
The coefficients to which are provided below.

\paragraph{$\widetilde{A}^{I,L}$ Coefficients:}
\begin{eqnarray}
\widetilde{a}_3^{I,L}&\approx&0\; ,\\
\widetilde{a}_2^{I,L}&\approx&-\frac{16\kappa_t}{x_By^2}\; ,\\
\widetilde{a}_1^{I,L}&\approx&\frac{8(y-2)\kappa_t^\frac{1}{2}[8M^2x_B^2-Q^2+t(4-6x_B)]}{Qx_By^2} \; ,\\
\widetilde{a}_0^{I,L}&\approx&\frac{8}{x_By^2} \bigg\{ M^2x_B^2(y^2-6y+6)+t[(y-2)^2-x_B(y^2-5y+5)] \bigg\}\; ,
\end{eqnarray}

\paragraph{$\widetilde{B}^{I,L}$ Coefficients:}

\begin{eqnarray}
\widetilde{b}_3^{I,L}&\approx&0\approx\widetilde{b}_2^{I,L}\;,\\
\widetilde{b}_1^{I,L}&\approx&-\frac{16tx_B(y-2)\kappa_t^\frac{1}{2}}{Q(x_B-2)y^2}\;,\\
\widetilde{b}_0^{I,L}&\approx&\frac{8tx_B(y-1)}{(x_B-2)y^2}\;,
\end{eqnarray}

\paragraph{$\widetilde{C}^{I,L}$ Coefficients:}

\begin{eqnarray}
\widetilde{c}_3^{I,L}&\approx&0\;,\\
\widetilde{c}_2^{I,L}&\approx&\frac{x_B}{x_B-2}\widetilde{a}_2^{I,L}\;,\\
\widetilde{c}_1^{I,L}&\approx&\frac{8(y-2)\kappa_t^\frac{1}{2}}{Q(x_B-2)^2y^2}\bigg\{8M^2x_B^2(x_B-2)-Q^2(x_B-2)-2t(4x_B^2-11x_B+7)\bigg\}\;,\\
\widetilde{c}_0^{I,L}&\approx&\frac{8(y^2-6y+6)\kappa_t}{(x_B-2)y^2(y-1)}\;.
\end{eqnarray}

\subsection{Unpolarized beam Transversely polarized Target}

We shall again decompose the $UT$ cross section here in terms of its \textit{in plane} and \textit{out of plane} components
\begin{equation}\label{sigIUT}
    \sigma^{UT}_\mathcal{I} =\cos(\phi_S-\phi)\sigma^{UT,\rm{in}}_\mathcal{I} + \sin(\phi_S-\phi)\sigma^{UT,\rm{out}}_\mathcal{I}\;.
\end{equation}

\paragraph{Parallel to Hadronic plane}

 \begin{equation}\label{sigIUTin}
     \sigma^{UT,\rm{in}}_\mathcal{I} = \frac{\Gamma}{Q^2t}(2\Lambda_T)F^{UT,\rm{in}}_\mathcal{I}\;,
 \end{equation}
 
\begin{eqnarray}\label{FIUTin}
 &F_{\mathcal{I}}^{UT,\rm{in}}=-\frac{2}{N} \Bigg\{\tilde A^{I,\rm{U}}\Bigg[\xi F_1\left(\xi \text{Im}\widetilde{ \mathcal {H}}+\left(\frac{\xi^2}{1+\xi}+\frac{t}{4M^2}\right)\text{Im}\widetilde{ \mathcal {E}} \right)+F_2\frac{t}{4M^2}\left((\xi^2-1)\text{Im}\widetilde{\mathcal {H}} +\xi^2\text{Im}\widetilde{ \mathcal {E}}\right)\Bigg]\nonumber\\
       &\qquad\qquad\qquad\qquad+\tilde B^{I,\rm{U}}(F_1+F_2)\left[\text{Im}\widetilde{ \mathcal {H}}+\left(\frac{t}{4M^2}-\frac{\xi}{1+\xi}\right)\xi\text{Im}\widetilde{\mathcal {E}} \right]\nonumber\\
       &\qquad\qquad\qquad\qquad+\tilde C^{I,\rm{U}} (F_1+F_2)\left[\xi \text{Im}\mathcal {H}+\left(\frac{\xi^2}{1+\xi}+\frac{t}{4M^2}\right) \text{Im}\mathcal {E} \right]  \Bigg\}\;.
\end{eqnarray}

\paragraph{Perpendicular to Hadronic plane}

 \begin{equation}\label{sigIUTout}
     \sigma^{UT,\rm{out}}_\mathcal{I} = \frac{\Gamma}{Q^2t}(2\Lambda_T)F^{UT,\rm{out}}_\mathcal{I}\;,
 \end{equation}
 
\begin{eqnarray}\label{FIUTout}
 &F_{\mathcal{I}}^{UT,\rm{out}}=\frac{2}{N} \text{Im}\Bigg\{A^{I,\rm{U}} \left[F_1\left(\xi^2\text{Im}\mathcal H+\left(\xi^2+\frac{t}{4M^2}\right)\text{Im}\mathcal E \right)+\frac{t}{4M^2}F_2\left((\xi^2-1)\text{Im}\mathcal H+\xi^2\text{Im}\mathcal E\right)\right]\nonumber\\
      &\qquad\qquad\qquad +B^{I,\rm{U}}(F_1+F_2) \left(\text{Im}\mathcal H+\frac{t}{4M^2}\text{Im}\mathcal E\right)-C^{I,\rm{U}} \xi(F_1+F_2)\left(\text{Im}\widetilde{\mathcal H}+\frac{t}{4M^2}\text{Im}\widetilde{\mathcal E}\right)\Bigg\}\;. \nonumber\\
\end{eqnarray}

\subsection{Polarized beam Transversely polarized Target}

Finally, the LT interference cross section is given by
\begin{equation}\label{sigILT}
    \sigma^{LT}_\mathcal{I} =\cos(\phi_S-\phi)\sigma^{LT,\rm{in}}_\mathcal{I} + \sin(\phi_S-\phi)\sigma^{LT,\rm{out}}_\mathcal{I}.
\end{equation}

\paragraph{Parallel to Hadronic plane}

 \begin{equation}\label{sigILTin}
     \sigma^{LT,\rm{in}}_\mathcal{I} = \frac{\Gamma}{Q^2t}(2h)(2\Lambda_T)F^{LT,\rm{in}}_\mathcal{I}\;,
 \end{equation}
 
\begin{eqnarray}\label{FILTin}
 F_{\mathcal{I}}^{LT,\rm{in}}&=& \frac{2}{N}  \Bigg\{\tilde A^{I,\rm{L}}\Bigg[\xi F_1\left(\xi \text{Re}\widetilde{ \mathcal {H}}+\left(\frac{\xi^2}{1+\xi}+\frac{t}{4M^2}\right)\text{Re}\widetilde{ \mathcal {E}} \right)+F_2\frac{t}{4M^2}\left((\xi^2-1) \text{Re}\widetilde{\mathcal {H}} +\xi^2\text{Re}\widetilde{ \mathcal {E}}\right)\Bigg] \nonumber\\
       &~&\qquad\qquad\qquad\qquad+\tilde B^{I,\rm{L}}(F_1+F_2)\left[\text{Re}\widetilde{ \mathcal {H}}+\left(\frac{t}{4M^2}-\frac{\xi}{1+\xi}\right)\xi \text{Re}\widetilde{\mathcal {E}} \right]\nonumber\\
       &~&\qquad\qquad\qquad\qquad+\tilde C^{I,\rm{L}} (F_1+F_2)\left[\xi \text{Re}\mathcal {H}+\left(\frac{\xi^2}{1+\xi}+\frac{t}{4M^2}\right)\text{Re} \mathcal {E} \right]  \Bigg\}\; .
\end{eqnarray}

\paragraph{Perpendicular to Hadronic plane}

 \begin{equation}\label{sigLTout}
     \sigma^{LT,\rm{out}}_\mathcal{I} = \frac{\Gamma}{Q^2t}(2h)(2\Lambda_T)F^{LT,\rm{out}}_\mathcal{I}\;,
 \end{equation}
 
\begin{eqnarray}\label{FILTout}
F_{\mathcal{I}}^{LT,\rm{out}}&=&-\frac{2}{N} \Bigg\{A^{I,\rm{L}} \left[F_1\left(\xi^2\text{Re}\mathcal H+\left(\xi^2+\frac{t}{4M^2}\right)\text{Re}\mathcal E \right)+\frac{t}{4M^2}F_2\left((\xi^2-1)\text{Re}\mathcal H+\xi^2\text{Re}\mathcal E\right)\right]\nonumber\\
     &~& \qquad\qquad \qquad+B^{I,\rm{L}}(F_1+F_2) \left(\text{Re}\mathcal H+\frac{t}{4M^2}\text{Re}\mathcal E\right)\nonumber\\
     &~&\qquad\qquad \qquad-C^{I,\rm{L}} \xi(F_1+F_2)\left(\text{Re}\widetilde{\mathcal H}+\frac{t}{4M^2}\text{Re}\widetilde{\mathcal E}\right)\Bigg\}\; .\nonumber\\
\end{eqnarray}

\section{Definition of Compton Form Factor Expressions}
\label{App:C}

We define here the recurring quadratic and linear expressions of CFFs which arise in our DVCS observables.  Starting with the quadratic expressions, we define
\begin{eqnarray}
\mathcal{D}_1^{\text{DVCS}}&=&(1-\xi^2)\big{[}(\text{Re}\mathcal{H})^2+(\text{Im}\mathcal{H})^2 +(\text{Re}\mathcal{\tilde{H}})^2+(\text{Im}\mathcal{\tilde{H}})^2 \big{]}\nonumber\\
    &~&-\frac{t}{4M^2}\big{[} (\text{Re}\mathcal{E})^2+(\text{Im}\mathcal{E})^2 +\xi^2\big{(}(\text{Re}\mathcal{\tilde{E}})^2+(\text{Im}\mathcal{\tilde{E}})^2 \big{)} \big{]}\nonumber\\
    &~&-\xi^2\Bigg{[}(\text{Re}\mathcal{E})^2+ (\text{Im}\mathcal{E})^2  +2\big{(}\text{Re}\mathcal{E}\text{Re}\mathcal{H} + \text{Im}\mathcal{E}\text{Im}\mathcal{H} \nonumber\\
    &~&+\text{Re}\mathcal{\tilde{E}}\text{Re}\mathcal{\tilde{H}} + \text{Im}\mathcal{\tilde{E}}\text{Im}\mathcal{\tilde{H}} \big{)}\Bigg{]}\;,
\end{eqnarray}

\begin{eqnarray}
\mathcal{D}_2^{\text{DVCS}}&=&  (1-\xi^2)(\text{Re}\mathcal{\widetilde{H}}\text{Re}\mathcal{H}+\text{Im}\mathcal{\widetilde{H}}\text{Im}\mathcal{H}) \nonumber\\
    &~&-\xi^2(\text{Re}\mathcal{\widetilde{H}}\text{Re}\mathcal{E}+\text{Im}\mathcal{\widetilde{H}}\text{Im}\mathcal{E}+\text{Re}\mathcal{\widetilde{E}}\text{Re}\mathcal{H}+\text{Im}\mathcal{\widetilde{E}}\text{Im}\mathcal{H}) \nonumber\\
    &~&-\bigg{(}\frac{\xi^2}{1+\xi}+\frac{t}{4M^2} \bigg{)}\xi (\text{Re}\mathcal{\widetilde{E}}\text{Re}\mathcal{E}+\text{Im}\mathcal{\widetilde{E}}\text{Im}\mathcal{E})\;,
\end{eqnarray}

\begin{eqnarray}
\mathcal{D}_3^{\text{DVCS}}&=&  \text{Re}\mathcal{H}\text{Im}\mathcal{E} - \text{Re}\mathcal{E}\text{Im}\mathcal{H} - \xi(\text{Re}\mathcal{\widetilde{H}}\text{Im}\mathcal{\widetilde{E}}-\text{Re}\mathcal{\widetilde{E}}\text{Im}\mathcal{\widetilde{H}})\;,
\end{eqnarray}

\begin{eqnarray}
\mathcal{D}_4^{\text{DVCS}}&=&\text{Re}\mathcal{\widetilde{H}}\text{Re}\mathcal{E} + \text{Im}\mathcal{E}\text{Im}\mathcal{\widetilde{H}} - \xi(\text{Re}\mathcal{\widetilde{E}}\text{Re}\mathcal{H}+\text{Im}\mathcal{\widetilde{E}}\text{Im}\mathcal{H}) \nonumber\\
&~&-\frac{\xi^2}{1+\xi}(\text{Re}\mathcal{\widetilde{E}}\text{Re}\mathcal{E}+\text{Im}\mathcal{\widetilde{E}}\text{Im}\mathcal{E})\;.
\end{eqnarray}
For the linear expressions which originate from the interference structure functions we have

\begin{eqnarray}
\mathcal{A}^U_{\rm{Re}, \rm{Im}}&=&F_1\begin{Bmatrix}\rm{Re} \\ \rm{Im} \end{Bmatrix}\mathcal{H}-\frac{t}{4M^2}F_2\begin{Bmatrix}\rm{Re} \\ \rm{Im} \end{Bmatrix}\mathcal{E}\;,\\
\mathcal{B}^U_{\rm{Re}, \rm{Im}}&=&(F_1+F_2)\bigg( \begin{Bmatrix}\rm{Re} \\ \rm{Im} \end{Bmatrix}\mathcal{H} + \begin{Bmatrix}\rm{Re} \\ \rm{Im} \end{Bmatrix}\mathcal{E} \bigg)\;,\\
\mathcal{C}^U_{\rm{Re}, \rm{Im}}&=&(F_1+F_2)\begin{Bmatrix}\rm{Re} \\ \rm{Im} \end{Bmatrix}\widetilde{\mathcal{H}}\;,
\end{eqnarray}
\begin{eqnarray}
\mathcal{A}^L_{\rm{Re}, \rm{Im}}&=&F_1\bigg( \begin{Bmatrix}\rm{Re} \\ \rm{Im} \end{Bmatrix}\widetilde{\mathcal{H}}-\frac{\xi^2}{1+\xi}\begin{Bmatrix}\rm{Re} \\ \rm{Im} \end{Bmatrix}\widetilde{\mathcal{E}}\bigg)-F_2\frac{t}{4M^2}\begin{Bmatrix}\rm{Re} \\ \rm{Im} \end{Bmatrix}\widetilde{\mathcal{E}}\;,\\
\mathcal{B}^L_{\rm{Re}, \rm{Im}}&=&(F_1+F_2)\bigg( \begin{Bmatrix}\rm{Re} \\ \rm{Im} \end{Bmatrix}\widetilde{\mathcal{H}}+\frac{\xi}{1+\xi}\begin{Bmatrix}\rm{Re} \\ \rm{Im} \end{Bmatrix}\widetilde{\mathcal{E}}\bigg)\;,\\
\mathcal{C}^L_{\rm{Re}, \rm{Im}}&=&-(F_1+F_2)\bigg( \begin{Bmatrix}\rm{Re} \\ \rm{Im} \end{Bmatrix}\mathcal{H}+\frac{\xi}{1+\xi}\begin{Bmatrix}\rm{Re} \\ \rm{Im} \end{Bmatrix} \mathcal{E}\bigg)\;,
\end{eqnarray}
\begin{eqnarray}
\mathcal{A}^{\rm{in}}_{\rm{Re}, \rm{Im}}&=&\xi F_1\Bigg(\xi\begin{Bmatrix}\rm{Re} \\ \rm{Im} \end{Bmatrix}\widetilde{\mathcal{H}}+\bigg(\frac{\xi^2}{1+\xi}+\frac{t}{4M^2} \bigg)\begin{Bmatrix}\rm{Re} \\ \rm{Im} \end{Bmatrix}\widetilde{\mathcal{E}} \Bigg) \nonumber\\
&~& +F_2\frac{t}{4M^2}\Bigg( (\xi^2-1)\begin{Bmatrix}\rm{Re} \\ \rm{Im} \end{Bmatrix}\widetilde{\mathcal{H}}+\xi^2\begin{Bmatrix}\rm{Re} \\ \rm{Im} \end{Bmatrix}\widetilde{\mathcal{E}} \Bigg)\;,\\
\mathcal{B}^{\rm{in}}_{\rm{Re}, \rm{Im}}&=& (F_1+F_2)\Bigg( \begin{Bmatrix}\rm{Re} \\ \rm{Im} \end{Bmatrix}\widetilde{\mathcal{H}}+\bigg(\frac{t}{4M^2}-\frac{\xi}{1+\xi} \bigg)\xi\begin{Bmatrix}\rm{Re} \\ \rm{Im} \end{Bmatrix}\widetilde{\mathcal{E}} \Bigg)\;,\\
\mathcal{C}^{\rm{in}}_{\rm{Re}, \rm{Im}}&=& (F_1+F_2)\Bigg(\xi \begin{Bmatrix}\rm{Re} \\ \rm{Im} \end{Bmatrix}\mathcal{H}+\bigg(\frac{\xi^2}{1+\xi}+\frac{t}{4M^2} \bigg)\begin{Bmatrix}\rm{Re} \\ \rm{Im} \end{Bmatrix}\mathcal{E} \Bigg)\;,
\end{eqnarray}
\begin{eqnarray}
\mathcal{A}^{\rm{out}}_{\rm{Re}, \rm{Im}}&=& F_1\Bigg(\xi^2\begin{Bmatrix}\rm{Re} \\ \rm{Im} \end{Bmatrix}\mathcal{H}+\bigg(\xi^2+\frac{t}{4M^2} \bigg)\begin{Bmatrix}\rm{Re} \\ \rm{Im} \end{Bmatrix}\mathcal{E} \Bigg) \nonumber\\
&~& +F_2\frac{t}{4M^2}\Bigg( (\xi^2-1)\begin{Bmatrix}\rm{Re} \\ \rm{Im} \end{Bmatrix}\mathcal{H}+\xi^2\begin{Bmatrix}\rm{Re} \\ \rm{Im} \end{Bmatrix}\mathcal{E} \Bigg)\;,\\
\mathcal{B}^{\rm{out}}_{\rm{Re}, \rm{Im}}&=& (F_1+F_2)\Bigg( \begin{Bmatrix}\rm{Re} \\ \rm{Im} \end{Bmatrix}\mathcal{H}+\frac{t}{4M^2}\xi\begin{Bmatrix}\rm{Re} \\ \rm{Im} \end{Bmatrix}\mathcal{E} \Bigg)\;,\\
\mathcal{C}^{\rm{out}}_{\rm{Re}, \rm{Im}}&=& (F_1+F_2)\xi\Bigg( \begin{Bmatrix}\rm{Re} \\ \rm{Im} \end{Bmatrix}\widetilde{\mathcal{H}}+\frac{t}{4M^2} \begin{Bmatrix}\rm{Re} \\ \rm{Im} \end{Bmatrix}\widetilde{\mathcal{E}} \Bigg)\;.
\end{eqnarray}

\section{Unpolarized Bethe-Heitler Cross Section}
\label{App:D}

The unpolarized Bethe-Heitler cross section can be expressed in terms of two kinematic coefficients, which can each be decomposed into 3 harmonic coefficients in the lab frame.  We express these coefficients explicitly in this appendix in the lab frame kinematics.
\begin{eqnarray}
\sigma^{UU}_{\text{BH}}&=&\frac{\Gamma}{\Omega_0 \mathcal{P}_1\mathcal{P}_2}\sum_{n=0}^2\Bigg[ a_n^{\text{BH}}\bigg( F_1^2-\frac{t}{4M^2}F_2^2 \bigg) +\frac{t}{2}b_n^{\text{BH}} G_M^2  \Bigg]\cos(n\phi)\;, 
\end{eqnarray}
where
\begin{equation}
    \Omega_0=-M^2x_B^2y^2Q^6(1+\gamma^2)^3\;.
\end{equation}
And each of the unapproximated harmonic coefficients are given by
\begin{eqnarray}
a_0^{\text{BH}}&=&8M^2Q^2(1+\gamma^2)\Bigg\{ Q^6\bigg[ 2M^2tx_B[-2x_B(y^2+y-1)+3y^2-4y+4] \nonumber\\
&~&\; +4M^4x_B^3(y^2+y-1) + t^2[-x_B(y-2)^2+y^2-2y+2]\bigg] \nonumber\\
&~&\; +Q^4\bigg[ M^2t^2x_B^2[-2x_B^2(y^2-2y+2)+4x_B(y^2+y-1)+y^2-2y+2]\nonumber\\
&~&\; -12M^4tx_B^4[(x_B-1)y^2+2y-2] +12M^6xB^6y^2\bigg] -Q^8(y^2-2y+2)\nonumber\\
&~&\; \times[t(xB-1)-M^2xB^2] -4M^4Q^2tx_B^4[2M^2xB^2y^2+t(xB^2-3x_B-1)y^2\nonumber\\
&~&\; -ty+t] +12M^6t^2x_B^6y^2\Bigg\}\;,
\end{eqnarray}

\begin{eqnarray}
a_1^{\text{BH}}&=&-16M^2x_BQ^5(y-2)(1+\gamma^2)\big[Q^4(t-2M^2x_B)+Q^2tx_B(t-2M^2(x_B-1))+2M^2t^2x_B^2\big]\nonumber\\
&~&\; \times\sqrt{1-y-\frac{y^2\gamma^2}{4}}\frac{\sqrt{Q^4[t(x_B-1)-M^2x_B^2]+Q^2tx_B[t(x_B-1)-2M^2x_B]-M^2t^2x_B^2}}{Q^2+tx_B}\;,\nonumber\\
\end{eqnarray}

\begin{eqnarray}
a_2^{\text{BH}}&=&32M^4x_B^2Q^2(1+\gamma^2)\bigg[Q^4(M^2x_B^2+t(1-x_B))+Q^2tx_B(2M^2x_B+t(1-x_B))+M^2t^2x_B^2 \bigg]\nonumber\\
&~&\;\times [M^2x_B^2y^2+Q^2(y-1)]\;,
\end{eqnarray}

 \begin{eqnarray}
 b_0^{\text{BH}}&=&4M^2txB^2Q^2(1+\gamma^2)\Bigg\{ Q^4\bigg[ -4M^2tx_B^2[(x_B+1)y^2-2y+2]+12M^4x_B^4y^2\nonumber\\
 &~&\; +t^2\big[2x_B^2(y^2-6y+6)-2x_B(y^2-6y+6)+y^2-2y+2 \big] \bigg]\nonumber\\
 &~&\; + 2Q^6\big[ 2M^2x_B^2(y^2+y-1)+tx_B(y^2-6y+6)-t(y-2)^2 \big]\nonumber\\
 &~&\; -4M^2Q^2tx_B^2\big[ 2M^2x_B^2y^2+t(x_B^2-x_B-1)y^2-ty+t \big] +12M^4t^2x_B^4y^2 +Q^8(y^2-2y+2)\Bigg\}\;,\nonumber\\
 \end{eqnarray}
 
 %
 \begin{eqnarray}
 b_1^{\text{BH}}&=&-\frac{tx_B[Q^2+t(2x_B-1)]}{Q^2(t-2M^2x_B)+2M^2tx_B}a_1^{BH}\;,
 \end{eqnarray}
 
 %
 \begin{eqnarray}
 b_2^{\text{BH}}&=&\frac{t}{2M^2}a_2^{BH}\;.
 \end{eqnarray}
 
 %
 %
 %

\section{Definition of Compound Coefficients}
\label{App:E}

When one multiplies a total DVCS cross section by the propagator factor $\mathcal{P}_1\mathcal{P}_2$, they are always left with a product of two harmonic series in the pure DVCS term.  This product can then be reduced into a single series by implementing multiple angle trigonometric identities.  We define one such recurring series' coefficients below.



Present in Eq.(\ref{sigUUlong}) are the following coefficients,

  \begin{eqnarray}
   (BH\otimes h^U)_0 &=& h_0^U(BH)_0+\frac{h_1^U(BH)_1}{2}+\frac{h_2^U(BH)_2}{2} \;,\\
   (BH\otimes h^U)_1 &=& h_0^U(BH)_1 +h_1^U(BH)_0 +\frac{h_1^U(BH)_2}{2} + \frac{h_2^U(BH)_1}{2} +\frac{h_3^U(BH)_2}{2} \;,\\
   (BH\otimes h^U)_2 &=& h_0^U(BH)_2 +h_2^U(BH)_0 +\frac{h_1^U(BH)_1}{2} + \frac{h_2^U(BH)_1}{2} \;,\\
   (BH\otimes h^U)_3 &=& h_3^U(BH)_0+\frac{h_1^U(BH)_2}{2}+\frac{h_2^U(BH)_1}{2} + h_3^U(BH)_0\;,
    \\
    (BH\otimes h^U)_4 &=&\frac{h_3^U(BH)_1}{2} + \frac{h_2^U(BH)_2}{2} + h_3^U(BH)_0  \;,\\
   (BH\otimes h^U)_5 &=& \frac{h_3^U(BH)_2}{2}\;.
     \end{eqnarray}





\bibliographystyle{jhep}
\bibliography{refs.bib}

\end{document}